\documentclass[onecolumn,authoryear,usenames,dvipsnames]{els-mrw}

\usepackage{BOONDOX-cal}
\usepackage[colorlinks=true, urlcolor=blue, linkcolor=red]{hyperref}
\usepackage{fancyvrb} 
\VerbatimFootnotes    
\newcommand{\caly}[1]{\ensuremath{\mathcal{#1}}}
\newcommand{\lum}{L}
\newcommand{\AMU}{\caly{A}} 
\newcommand{\NUM}{\caly{N}} 
\newcommand{\Zch}{\caly{Z}} 
\newcommand{\period}{\caly{P}} 
\newcommand{\prob}{\caly{p~}} 
\newcommand{\epsim}{\caly{\epsilon}} 
\newcommand{\rat}{\caly{r}} 

\newcommand{\lsun}{\mathrm{L}_{\odot}}
\newcommand{\rsun}{\mathrm{R}_{\odot}}

\newcommand{\msun}{\mathrm{M}_{\odot}}

\newcommand{\teff}{{T}_{\text{eff}}}

\newcommand{\mini}{\textit{M}_{\text{ini}}}
\newcommand{\mhef}{\textit{M}_{\text{Flash}}}
\newcommand{\mhe}{\textit{M}_{\text{He}}}
\newcommand{\mco}{\textit{M}_{\text{CO}}}
\newcommand{\mch}{\textit{M}_{\text{Ch}}}
\newcommand{\magb}{\textit{M}_{\text{AGB}}}
\newcommand{\msagb}{\textit{M}_{\text{SAGB}}}

\newcommand{\kros}{{\kappa}_\mathrm{R}}

\newcommand{\stefb}{$\sigma_{\text{SB}}$}
\newcommand{\X}{\textit{X}}
\newcommand{\Y}{\textit{Y}}
\newcommand{\Z}{\textit{Z}}
\newcommand{\Ai}{\textit{A}_{\mathrm{i}}}

\newcommand{\kb}{\text{k}_{\mathrm{B}}}
\newcommand{\G}{\text{G}}
\newcommand{\su}[1] {\sum_{\mathrm{#1}}}

\newcommand {\reac}[6] {$\rm\,{}^{#2}\kern-0.8pt{#1}\,+{#3}\,\rightarrow {#4}+
  \,{}^{#6}\kern-0.8pt{#5}\,$}
\def\parsec/{\textsc{parsec}}
\def\aesopus/{\textsc{\ae sopus}}
\def\atomic/{\textsc{atomic}}
\def\emcee/{\texttt{emcee}}
\newcommand{\eq}[1]{Equation~({#1})} 

\newcommand{\araa}{Annu. Rev. Astron. Astrophys.} 
\newcommand{\aj}{Astron. J.} 
\newcommand{\apj}{Astrophys. J.} 
\newcommand{\apjl}{Astrophys. J. Lett.} 
\newcommand{\apjs}{Astrophys. J. Suppl. Ser.} 
\newcommand{\apss}{Astrophys. Space Sci.} 
\newcommand{\aap}{Astron. Astrophys.} 
\newcommand{\jgr}{J. Geophys. Res.} 
\newcommand{\mnras}{Mon. Not. R. Astron. Soc.} 
\newcommand{\nat}{Nature} 
\newcommand{\physrep}{Phys. Rep.} 
\newcommand{\prd}{Phys. Rev. D} 
\newcommand{\pasj}{Publ. Astron. Soc. Jpn} 
\newcommand{\zap}{Zeitschrift f\"ur Astrophysik} %

\newcommand{\pdv}[2]{  \frac{\partial{{#1}}}{\partial{{#2}}}}
\newcommand{\pdvh}[2]{  \partial{{#1}}/ \partial{{#2}}}
\newcommand{\diff}[2]{\frac{{d}{#1}}{{d}{#2}}}
\usepackage{amsmath,amssymb,amsfonts,amsthm,makeidx,graphicx}
\usepackage{txfonts}
\usepackage{helvet}
\usepackage{calligra}
\usepackage{floatrow}
\DeclareMathAlphabet{\mathcalligra}{T1}{calligra}{m}{n}
\DeclareFontShape{T1}{calligra}{m}{n}{<->s*[1.3]callig15}{}
\usepackage{dirtytalk}

\usepackage{booktabs}

\usepackage{color}
\begin{document}

\chapter{Evolution and final fates of low- and intermediate-mass stars}\label{chap1}

\author[1]{Alessandro Bressan}%
\author[1]{Kendall Gale Shepherd}%


\address[1]{\orgname{SISSA}, \orgdiv{Physics Department}, \orgaddress{Via Bonomea 265, Trieste Italy}}

\articletag{Chapter Article tagline: update of previous edition,, reprint..}

\maketitle

\begin{abstract}[Abstract]
Stars are unique bodies of the Universe where self-gravity compress matter to such high temperature and density that several nuclear fusion  reactions ignite, providing enough feedback against further compression for a time that can be even larger than the age of the universe. The main property of a star is its mass because it determines its structure, evolutionary history, age, and ultimate fate. Depending on this quantity, stars are broadly classified as low-mass stars, like our Sun,  intermediate mass stars as the variable star Delta Cephei, and massive stars as  Betelgeuse, a red supergiant star in Orion constellation.
Here we will introduce the basic notions useful to understand stellar evolution of low- and intermediate- mass stars. {This mass range (0.1 $\msun -  10.0 ~\msun$) deserves special attention, as it contains most of the stars in the universe. 
This chapter will focus on how these stars form, the processes that drive their evolution, and key details regarding their structure. Finally, we will discuss the death of such stars, emphasizing the unique fates associated with low- and intermediate-mass stars. } 
\end{abstract}



\begin{glossary}[Glossary]
\end{glossary}

\begin{glossary}[Nomenclature]
\begin{tabular}{@{}lp{34pc}@{}}
AGB &Asymptotic Giant Branch - section~\ref{secagb}\\
CMD &The Colour Magnitude Diagram - section~\ref{subsecHRD}\\
EOS &The equation of state of stellar plasma - section~\ref{chap1:subsub_mass}\\
HB &Horizontal Branch - section~\ref{secHB}\\
HRD &The Hertzsprung-Russell Diagram - section~\ref{subsecHRD}\\
MLT &Mixing Length Theory of convection - section~\ref{secconv}\\ 
MS & Main Sequence - section~\ref{secmainseq}\\
PAGB &Post Asymptotic Giant Branch - section~\ref{secpagb}\\
RGB &Red Giant Branch - section~\ref{secrg}\\
SAGB &Super-Asymptotic Giant Branch - section~\ref{secagb}\\
SGB &Sub Giant Branch - section~\ref{chap2:subsec2}\\
TRGB &The tip of the RGB - section~\ref{chap1:Heflash}\\
WD &White Dwarf - section~\ref{secwd}\\
ZAMS &Zero Age Main Sequence - section~\ref{secmainseq}\\
\end{tabular}
\end{glossary}

\section{Objectives}\label{seckey}
\begin{itemize}%
\item The system of equations of the stellar structure
\item Chemical composition of stars
\item The Hertzsprung-Russell Diagram
\item Hydrogen burning- and Helium burning- nuclear reactions
\item Main sequence stars
\item Stellar lifetimes
\item Sub giant stars
\item Electron degeneracy
\item Red Giant stars
\item Horizontal Branch stars
\item Asymptotic Giant Branch stars
\item White Dwarfs
\end{itemize}

\section{Introduction}\label{chap1:sec1}
In this chapter we will describe the evolution of low- and intermediate-~mass stars, namely stars whose initial mass is from one tenth to about ten times that of the Sun. Section \ref{secstruct} will deal with the basic but general equations that govern the stellar structure, essentially the mass, momentum and energy conservation, and the equation of energy transport. The method of solution will also be briefly discussed, and a list of existing numerical codes adopted by different authors will be provided. In the following sections we will describe the evolution of low-mass stars
during the Pre-Main Sequence phase (section~\ref{pre-main-seq}) that is, just after their formation;
the Main Sequence phase (section~\ref{secmainseq}), where stars burn their central Hydrogen (H);
the Red Giant phase with the Helium (He) Flash at He ignition (section~\ref{secrg});
the Horizontal Branch (section~\ref{secHB})
 where low mass stars burn their central He. Section~\ref{secims} will briefly review the evolution of intermediate mass stars until central He exhaustion while, the  
final nuclear phases of low- and intermediate-mass stars, being very similar, will be discussed together in section~\ref{secfinal}.
In section \ref{secwd}, we will discuss how the final fate of such stars is approached, without knowing if an end really exists.  
This chapter is meant to provide a broad overview on low- and intermediate-mass stars, and therefore some topics are treated concisely for the sake of brevity. We refer the reader to specific textbooks cited in the text for more details.

\section{The structure of stars}\label{secstruct}
While discussing stellar structure and evolution it is unavoidable to refer to the Sun. A few preliminary considerations concerning the Sun are useful to introduce the reader to the realm of stars.
The Sun is the star nearest to us; the Earth orbits around the Sun at an average distance of 150 million km in a period of time called a year, which corresponds to 3.1 $\times$ 10$^7$ seconds, or about 365.25 days. Nevertheless the Sun's luminosity is so high that the heat captured by the Earth  helps life to develop and evolve on its surface. 
{Its age is estimated to be about 4.5 billion years.}
Its distance from the other nearest star, Proxima Centauri, is about 4.23 light-years \footnote{A light-year (ly) corresponds to 9.461$\times$10$^{17}$cm; 1 parsec is 3.26156 ly.}.  Therefore, the Sun is an isolated star, and its structure and evolution depend only on its intrinsic properties\footnote{Notable exceptions are tight binaries or other multiple stellar systems where individual components feel the effects of the nearby companions.}.

Evidence suggests that the geometry of the Sun can be very well approximated by a sphere. All other stars are so distant that we see them as point-like sources\footnote{Only a few other stars are sufficiently near to us that we may obtain an extended image, i.e. a non point-like image.}.
Besides the evidence, spherical symmetry is also prompted by self-gravity and it is usually a good approximation. 
In a spherically symmetric model all physical quantities depend only on the radial coordinate, so that a three dimensional real star can be described with an one dimensional model. {The benefits of using a one dimensional model will become apparent in the following sections.} However  deviations from spherical symmetry may arise because of non-central forces, like those originating from rotation and magnetic fields. These effects will not be considered here \cite[but see e.g.][]{Maeder2009,Low01}.   

We will now derive the equations that regulate the stellar structure assuming a spherical configuration.  We first define the following basic quantities at a given radial distance from the center, $r$. The mass enclosed within the sphere of radius ${r}$ is ${m(r)}$, ${\rho(r)}$ the matter density, ${P(r)}$ the pressure, ${T(r)}$  the temperature, $L(r)$  the local luminosity (the energy flowing through the surface of radius ${r}$ per unit time), $\mathrm{\epsilon_{nuc}(r)}$ is the energy generated by nuclear reactions per unit mass and time, $\mathrm{\epsilon_\nu(r)}$ is the energy lost in form of neutrinos per unit mass and time, ${s(r)}$ is the specific entropy (i.e. the entropy per unit mass). 
All of the above quantities are constant on the surface of a sphere of radius ${r}$, and the problem of defining the structure of a star is that of describing
their variation over a radial displacement between ${r}$ and a nearby location ${r+dr}$ (differential equations), plus that of finding their initial values, the so called initial conditions.
\subsection{The mass conservation}\label{chap1:subsub_mass}
In the case of the mass enclosed by the spherical surface at ${r}$, ${m(r)}$, its variation is simply the mass contained by the spherical shell limited by ${r}$ and ${r+dr}$, whose volume is ${dV=4 \pi r^2 dr}$. Thus the first {differential equation} of the stellar structure, stating the principle of mass conservation, writes \citep{Kippenhahn2013,Maeder2009}
\begin{equation}
    {\pdv{m}{r}=4\pi r^2 \rho}~.\label{eq:mass_conservation}
\end{equation}
This differential equation involves two dependent functions of the independent variable ${r}$, ${m(r)}$ and  ${\rho(r)}$ and at least another differential equation involving the same functions must be found to get a viable solution.
We note that the matter density ${\rho(r)}$ is related to the pressure ${P(r)}$ and  the temperature  ${T(r)}$  through the equation of state (EOS).
\subsection{The hydrostatic equilibrium}\label{chap1:subsub_hydro}
A differential equation for the internal pressure can be obtained by considering that the Sun has maintained its size almost unchanged for at least centuries. This indicates that the attraction by  gravity at any point in its interior is  balanced by another force, otherwise the Sun would collapse in a  {dynamical timescale}, which is less than a fraction of a hour. We say that the Sun satisfies the conditions for the hydrostatic equilibrium, with the balancing force being provided by the {pressure gradient}, $\pdvh{P}{r}$.
Equating the outward acceleration due to the radial pressure gradient and the inward acceleration due to gravity we get
\begin{equation}
 { -\frac{1}{\rho}\pdv{P}{r}-\frac{\G m}{r^2}=0}\label{eq:hydro}~.
\end{equation}
 Note that the quantity ${\pdvh{P}{r}}$ is negative so that the first term in the left hand side of \eq{\ref{eq:hydro}} is positive and directed  outward while
 the acceleration of gravity is negative and directed inward (opposite of the radial vector).
\begin{BoxTypeA}[chap1:box1]{Hydrostatic Equilibrium}
\section*{Central pressure and central temperature}
\eq{\ref{eq:mass_conservation}} and \eq{\ref{eq:hydro}} can be combined to give \citep[see e.g.][]{Kippenhahn2013}
\begin{equation}
  {\pdv{P}{m}=-\frac{\G m}{4\pi r^4}}\label{eq:hydro_inmass} ~.
\end{equation}
From \eq{\ref{eq:hydro_inmass}} we can get an order of magnitude estimate for central 
pressure
\begin{align}
    \frac{P_{\mathrm{central}}-P_{\mathrm{surface}}}{m_{\mathrm{central}}-m_{\mathrm{surface}}}&{\simeq-\frac{\G M}{4\pi R^4}}\\  \nonumber
 P\mathrm{_{central}}\simeq\frac{\G M^2}{4\pi R^4}&\mathrm{\simeq\frac{6.67~10^{-8} (2 \times 10^{33})^2}{4 \pi (7\times 10^{10})^4}dyn~cm^{-2} \simeq 10^{15}dyn~cm^{-2}} ~,
\end{align}
where we have assumed for the Sun  $\mathrm{M_\odot\sim~2\times10^{33}~g}$ and $\mathrm{R_\odot\sim~7\times10^{10}~cm}$  together with $M\mathrm{_{central}}$=0 and $P\mathrm{_{surface}}$=0. 

If $M$ and $V$ are the mass and the volume of the gas, and $\NUM$ the number of atoms  such that, e.g. for pure H,  $M=\NUM{m_\mathrm{H}}$, the perfect gas law equation of state is $P~V = \NUM$ $\kb T$ . Since
$\NUM=M/m\mathrm{_H}$ and noting that 
$\rho=M/V$ we can write
\begin{equation}
{P}~{ =\frac{\rho}{m_\mathrm{H}} \kb T}~.\\
\label{eq:eos}
\end{equation}
Then, using for the density $\rho$ its average value in the Sun $\mathrm{M_\odot/(\frac{4}{3}~\pi~R_\odot^3)\simeq~1g~cm^{-3}}$
we get an order of magnitude estimate of the central temperature
\begin{align}
    T\mathrm{_{central}}&{\simeq 10^{15}\frac{m_\mathrm{H}}{\rho \kb} \simeq 1.2 \times 10^7 \text{K}}~.
\end{align}

\subsection*{The dynamical timescale.}
In case the condition for the  hydrostatic equilibrium is violated, an interior mass element is subject to a residual acceleration
\begin{equation}
   {\pdv{^2r}{t^2}=-\frac{1}{\rho}\pdv{P}{r}-\frac{\G m}{r^2}},\label{eq:hydro_acc}~.
\end{equation}
 Assuming that the unbalance between gravity and pressure gradient is a fraction $f$ of the gravity, then
\begin{equation}
   \pdv{^2r}{t^2}={f \frac{\G m}{r^2}} \label{eq:hydro_acc_umbal}~,
\end{equation}
and the corresponding dynamical timescale would be \begin{equation}
 {t_\mathrm{d}^2=\frac{1}{f \G \frac{m}{r^3}}\simeq\frac{1}{4 f \G <\rho(r)>}}~.
\label{eq:hydro_timescale}
\end{equation}
Again, using the average density of the Sun and assuming $f\sim$1, $t_\mathrm{d}\sim~30$~min, meaning that if the Sun were not in hydrostatic equilibrium, it should appreciably change it radius in a timescale of about 30 min \citep[e.g.][]{Kippenhahn2013}.
Conversely, the slowness with which the Sun is actually changing its radius suggests that it follows the hydrostatic equilibrium to an extremely great
precision. Indeed, accurate time domain observations from {helioseismology} indicate that the Sun 
oscillates at frequencies that are in agreement with \eq{\ref{eq:hydro_timescale}} \citep[][and references therein]{Buldgen2019}.
\end{BoxTypeA}
\subsection{The temperature gradient}\label{chap1:subsub_etrans}
In general, \eq{\ref{eq:mass_conservation}} and \eq{\ref{eq:hydro}} are not enough to close the system of differential equations of the stellar structure  because
they involve three dependent functions of the independent variable ${r}$\footnote{A favorable condition arises when the pressure is a function of only the density. In that case the structure can be defined by the two equations (\ref{eq:mass_conservation}) and (\ref{eq:hydro}) if suitable boundary conditions are found, because they involve only two dependent functions of radius, the pressure $P(r)$ and  and the mass $m(r)$.}. Using the equation of state will generally introduce another dependent function of radius, the temperature $T$, and a new differential equation for $T(r)$ must be introduced.
The equation for the change of the temperature with radius is strictly connected with the mechanism responsible for the energy transport within stars. 

From laboratory experiments we know that there are at least three main processes to transport energy within a medium. One is  {conduction} by free electrons, the other is {convection} by moving fluid elements, and the third one is  {radiation} by photons. Within stars the main mechanisms are radiation and  convection, while conduction becomes important only in particular conditions. 

\begin{BoxTypeA}[chap3.3:box1]{The mean free path of photons in the interiors of our Sun is of the order of a few centimeters}\\
{The mean free path refers to the average distance travelled by a particle in between collisions.}
For photons, we get it by multiplying  the average time between collisions ($\mathrm{\tau_{coll}}$) of photons with  electrons, by the velocity of the photons.
$\mathrm{\tau_{coll}}$ Is the reciprocal of the number of collisions per unit time, $N\mathrm{_{coll}}$. $N\mathrm{_{coll}}$ Is the number of possible collisions with electrons within the effective volume sweep up by a moving cross-section of the photon-electron interaction, per unit time. For example, for the  {Thomson scattering} of photons by free electrons in an ionized plasma, $N\mathrm{_{coll}}$ is the product of the number density of electrons ($n\mathrm{_e}$) times the volume of the cylinder with base   $\mathrm{\sigma_T}$ and height $h = v\mathrm{_{phot}*1 sec}$: $N\mathrm{_{coll}}=n_\mathrm{e}~\sigma_\mathrm{T}~v_{\mathrm{phot}}$. Thus the mean free path $\mathrm{\ell_{phot}}$ is
\begin{equation}
    \ell_{\mathrm{phot}}=v_{\mathrm{phot}}~\tau_{\mathrm{coll}}= \frac{v_{\mathrm{phot}}}{N_{\mathrm{coll}}}=\frac{v_{\mathrm{phot}}}{n_\mathrm{e}~\sigma_\mathrm{T}~v_{\mathrm{phot}}}=\frac{1}{n_\mathrm{e}~\sigma_\mathrm{T}} ~.
\end{equation}
The mean free path of a particle for a certain interaction does  not depend on  the relative velocity of the colliding particles. The cross section of the Thomson scattering is $\mathrm{\sigma_T\simeq6.6525~10^{-25}cm^2}$ and $n\mathrm{_e}\simeq\rho/m_\mathrm{H}$  cm$^{-3}$. For the interior of the Sun,   $\rho\geq$1g~cm$^{-3}$ and with $m_\mathrm{H}$=1.6735 $\times~10^{-24}$g we get
\begin{equation}
    \ell_\mathrm{phot}=\frac{1}{n_\mathrm{e}~\sigma_\mathrm{T}}=\frac{m_\mathrm{H}}{\rho~\sigma_\mathrm{T}}\leq\frac{1.67~\times~10^{-24}}{1.\times6.65 \times 10^{-25}} \simeq 2.5 \mathrm{cm} ~.
\label{photmeanfp}
\end{equation}
Thus, photons within the Sun travel, on average, a few centimeters or even less between each interaction.
\end{BoxTypeA}

\begin{BoxTypeA}[chap3.3:box2]{The stellar radius and the effective temperature}\\
Making use of the results of the  theory of stellar atmospheres \cite[e.g.][]{Mihalas1970} the stellar radius is conventionally defined as the radial coordinate where the optical depth of photons against interaction with the stellar plasma, beginning from zero at infinity, reaches 2/3 as the stellar interiors are approached:
$\tau_\mathrm{phot}=-\int_\infty^R\kros\rho{dr}=2/3$.
The star extends beyond that radius and formally up to infinity, but with a density that becomes so low that it has no significant impact on the star.
Within this external region of the star, where the photons can freely escape because $\ell_\mathrm{phot}/R>>1$, the medium is said to be optically thin. In deeper regions $\tau$ increases and the region is said to be optically thick, being $\ell_\mathrm{phot}/R<<1$. The thermal stellar radiation that hits our detectors thus carries information only from this narrow  region where $\tau\sim 2/3$, called the photosphere. 

With the luminosity and the radius, an effective temperature, $\teff$, is defined by  assuming that the emissivity at the photosphere  equals that of a black body. By applying the Stefan–Boltzmann law, one gets
\begin{equation}
    \frac{L}{4\pi{R^2}}=\sigma_{\mathrm{SB}}{\teff^4}~,
\label{eqteff}
\end{equation}
where \stefb\ = ~6.670374419~$\times$~10$^{-5}$~erg/s/cm$^2$/K$^4$ is  the Stefan-Boltzmann constant.
The effective temperature can be measured in several different ways \citep[see e.g.][]{Smalley05,Gray19}.
The nominal value for the Sun is 
${\teff}_{\odot}$=5772~K \citep{prsa16}.
\end{BoxTypeA}

The energy transport mechanisms are responsible of keeping different stellar layers in thermal 
contact and they involve the knowledge of the  {mean free path}, $\ell$, of the carriers of energy, i.e. the average distance  the carriers travel from the site where they 
absorb heat to the site where they deposit heat. This average distance depends on the particular ability of the carrier to interact with the surrounding medium,  the  {cross section 
$\mathrm{\sigma_{interaction}}$ of the interaction. In the case of conduction, a free electron acquires energy in 
the collision with another charged particle and travels a certain distance before the following interaction, where an exchange of energy may  occur. The same is true for photons that 
are emitted in a hotter region and then, after traveling a certain distance, are absorbed by particles in a cooler region.  
Convection develops in a medium within a gravity field when hotter (and less dense) macroscopic bubbles move because of the Archimedes buoyant force. 
In that case the mean free path of the bubbles is called the mixing length, because it is supposed that bubbles keep their identity until 
they mix with the surrounding medium \citep[e.g.][]{Bohm-Vitense}. Generally, the larger the mean free path of a carrier, the more efficient it is in transporting energy.
For photons and electrons, the mean free path is calculated by considering the average distance traveled between two subsequent collisions.\\

\begin{BoxTypeA}[chap3.3:box3]{Adjacent layers of matter within the Sun are in thermal equilibrium}

Adjacent means that the layers are at a distance of a photon mean free path so that radiation may keep them in thermal contact.
For the central temperature of the Sun, the estimate derived from  the hydrostatic equilibrium equation gives $T\mathrm{_{central}\simeq~ 10^7K}$. The average temperature gradient within the Sun, $dT$/$dr$, is then
\begin{equation}
    \frac{T_{\mathrm{central}}-T_{\mathrm{surface}}} {R_{\mathrm{central}}-R_{\mathrm{surface}}}\sim-  \frac{T_{\mathrm{central}}}{R_{\mathrm{surface}}} =
    - \frac{10^7}{7 \times 10^{10}}\simeq~-1.4 \times 10^{-4} \mathrm{K~cm^{-1}}~.
\end{equation}

The temperature variation over a photon mean free path is ${{\delta}T\simeq~d}T/dr~\mathrm{\times~\ell\simeq~1.4 \times 10^{-4}\times2.5}$=~3.5 $\times$ 10$^{-4}$ K.
What matters is the relative temperature variation, which is ${{\delta}T/T}\simeq~$3.5 $\times$ 10$^{-4}$/$10^7$.
\noindent So, the relative temperature variation between shells of matter kept in contact by photons is ${{\delta}T/T}\simeq$ ~3.5 $\times$ 10$^{-11}$. Thus, internal layers are \say{locally} in thermal equilibrium (their relative temperature difference is negligibly small). Since the layers are also in hydrostatic equilibrium, we say that they are in Local Thermodynamic Equilibrium (LTE). The star, as a whole, is not in thermodynamic equilibrium because it is an open system that is losing energy through photons, neutrinos and even mass flows.
\end{BoxTypeA}


\subsubsection{Radiative energy transport}\label{chap1:subsub_radtrans}
The relative short length of the photon mean free path, \eq{\ref{photmeanfp}},  with respect to the radius of the Sun suggests that the energy transport by photons is a diffusive process. In this case the density of the energy flux, $\lum/4 \pi r^2$, 
is proportional to the gradient of the energy density of radiation, $U=\mathrm{a}T^4$,
where $\mathrm{a}$ = 4\stefb/c,   is the radiation constant.
The constant of proportionality is the diffusion coefficient which has the form
\begin{equation}
    {D}=\frac{1}{3} \mathrm{c} \ell ~,\label{eq:diff_coef}~,
\end{equation}
where c is the speed of light and $\ell$ is the  photon mean free path. The density of the energy flux for the radiative transport is thus
\begin{equation}
\frac{\lum}{4 \pi {r}^2}=-\frac{1}{3}\mathrm{c}\frac{1}{n\mathrm{_e} \sigma \mathrm{_T}}\nabla{U} = -\frac{4\mathrm{ac}}{3}\frac{{T}^3}{n\mathrm{_e} \sigma\mathrm{_T}}\diff{T}{r} \nonumber~.
\label{eq:rtransport1}
\end{equation}
Inverting this equation, we get the temperature gradient equation for the radiative transport
\begin{equation}
\diff{T}{r}=-\frac{3 n\mathrm{_e} \sigma_\mathrm{T}}{4\mathrm{ac}T^3} \frac{{\lum}}{4 \pi {r}^2}~.\\
\label{eq:rtransport}
\end{equation}
In \eq{\ref{eq:rtransport}} we made use of the microscopic Thomson cross section for the scattering of photons by free non relativistic electrons. With more generality we may consider other processes, including interactions of photons with atomic bound electrons (bound-bound interactions), photo-ionization (bound-free interactions), interactions with free electrons near charged particles (free-free), Compton (for energetic photons) scattering and also interactions with molecules. These interactions depend on the frequency of the photons and to include them into \eq{\ref{eq:rtransport}}, one must consider a suitable average over the whole frequency range of radiation. It has been shown that a suitable mean free path for photons can be obtained using the Rosseland mean opacity of stellar matter $\kros$ [cm$^2$g$^{-1}$] \citep[][]{Rosseland1924}
\begin{equation}
\ell=\frac{1}{\kros \times \rho} ~,
\end{equation}
where $\kros$ is a function of temperature, density and elemental abundance of a specific stellar layer, i.e.
$\kros=\kros{[T(r),\rho(r), \X_\mathrm{i}(r)]}$.  
Here and later, the term $\X\mathrm{_i}$  indicates the abundance by mass fraction of all the elements that are explicitly considered in the calculation of a stellar structure.

Using the Rosseland mean opacity, which is computed in several laboratories,
the temperature gradient equation for the radiative transport can be written as  \citep[e.g.][]{Weiss2004}\footnote{Because energy transport by electron conduction can also be treated as diffusive processes, conduction is usually accounted for by calculating a \say{conductive opacity}, $\mathrm{K_C}$, and applying the following correction to the Rosseland mean opacity: $\mathrm{1/\kappa_{tot}=1/\kros+1/\kappa_C}$ \citep[e.g.][]{Cassisi21}.}
\begin{equation}
\pdv{T}{r}=\pdv{T}{r}\Bigr\rvert_{\mathrm{Rad}}={-\frac{3 \kros \rho}{4\mathrm{ac}T^3} \frac{\lum}{4 \pi r^2}}~.\\
\label{eq:rtransportross}
\end{equation}
\subsubsection{Convective energy transport}
\label{secconv}
\eq{\ref {eq:rtransportross}} states that energy that diffuses outward by radiation establishes a local negative temperature gradient whose absolute value increases with increasing opacity ($\kros$)  or with increasing flux density (${\lum/4\pi r^2}$). 
Laboratory experiments on convection show that, if the absolute value of the local temperature gradient exceeds the adiabatic one, the medium become unstable and convection sets in, becoming the most efficient energy transport mechanism. 
Making use of the logarithmic temperature versus pressure gradient $\nabla = (d\mathrm{ln}T / ~d\mathrm{ln}P) = (P/T)(dT/dP)$,
the condition to adopt either the radiative or the adiabatic  gradient in  \eq{\ref{eq:rtransportross}} is \citep[see e.g.][]{Kippenhahn2013}\footnote{Note that $\nabla$ has the opposite sign of $\pdv{T}{r}$.}:  
\begin{align}
    \nabla=&\mathrm{\nabla_{Rad} \quad if \quad  \nabla_{Rad} < \nabla_{Adi} \quad (stable)}\nonumber \\
    \nabla=&\mathrm{\nabla_{Adi} \quad if \quad   \nabla_{Rad} \geq \nabla_{Adi}  \quad (unstable)}~.
    \label{eq:convstab}
\end{align}

In \eq{\ref{eq:convstab}}, 
$\nabla \mathrm{_{Rad}}$ and $\mathrm{\nabla_{Adi}}$ are the values of the radiative gradient needed to transport the luminosity calculated with \eq{\ref{eq:rtransportross}}, and  the gradient
of a quasi-static thermodynamic adiabatic process, respectively. Both $\nabla \mathrm{_{Rad}}$ and $\mathrm{\nabla_{Adi}}$  can be calculated from the local values of ${T}$, $\rho$, $\X\mathrm{_i}$, of the structure. 
Equation (\ref{eq:convstab})  states the  Schwarzschild criterion for stability against convection (named after Karl Schwarzschild). Since a negative gradient of mean molecular weight favors stability, a more stringent condition is often used, the Ledoux one (named after Paul Ledoux)\footnote{Both the Schwarzschild and the Ledoux criteria define the stability of the medium against convection, but there remains an ambiguity on the real size of convective regions because convection may penetrate into the surrounding stable regions, a process named {\sl convective overshooting} \citep[e.g.][]{Bressan1981}.}. 

{The Suns' convective zones produce currents of plasma, that cause the appearance of granules at the top of the convective cells.}
The Solar granulation observed at the Suns' photosphere \citep{Bahng1961}, is therefore indicative of the presence of convective currents. Helioseismology shows that solar convection extends from the surface down to the 71.3\% of the solar radius \citep{BasuAntia97}. Below that point, radiation is able to transport the energy flux without destabilizing the plasma.  The geometric size of these convective cells offer useful indicators of the scale of this process, which cannot be determined from first physical principles.

In the more external regions of a star, convective energy transport may become inefficient due to the low density of the medium. Though very thin in mass, these regions may contribute to a significant fraction of the stellar radius (because of the low density), and so they may have a significant impact on  the effective temperature of the star.
In this case, convective elements do not move adiabatically and their temperature gradient must be derived by explicitly solving the equations of the Mixing Length Theory (MLT) of convection \citep[e.g.][]{Bohm-Vitense, Weiss2004}. This theory is based on a typical scale for convective motions, the Mixing Length (ML). 
The ML is one of the more important but still uncertain parameters in stellar evolution, and it is usually determined by comparison of suitable solar models with the real Sun. 

\subsection{The energy conservation}
The differential \eq{\ref{eq:rtransportross}} provides the variation of the temperature within the star, but the whole system of equations cannot be integrated because \eq{\ref{eq:rtransportross}} introduces the function $\lum(r)$, the luminosity, which is still unknown.
The last term in the RHS of \eq{\ref{eq:rtransportross}} describes the amount of energy that flows per unit time and per unit area across the spherical surface at ${r}$. If there are no sources or sinks of energy, the total ${\lum(r)}$ over the whole surface, $4\pi r^2$, will be constant. 
Conversely, if sources and sinks are present in the mass shell ${dm=4\pi r^2 \rho dr}$, and we quantify them as  
$\epsim$ [erg~s$^{-1}$g$^{-1}$], the luminosity will change by  ${\epsim\times{dm}}$, so that we may write the energy conservation as
\begin{equation}
\pdv{\lum}{r}={4\pi~r^2\rho\epsim} ~.
\label{eq:luminosity}\\
\end{equation}
The main sources or sinks of energy
within stars are due to nuclear reactions, work done by gravity or by pressure against gravity (contraction or expansion of stellar layers) and processes that involve neutrino production that, because of their negligible cross section with matter, escape directly from the star instead of diffusing outside like photons\footnote{An exception is the formation of a neutrino-sphere during the final collapse of massive stars \citep[e.g.][]{Burrows21}.}.
These terms are explicitly indicated as
\begin{equation}
    \epsim(r)=\epsilon\mathrm{_{nuc}}(r)-T(r)\pdv{s(r,t)}{t}-\epsilon_\nu(r) ~,
\label{eq:epsim}
\end{equation}
where $\mathrm{\epsilon_{nuc}}(T,\rho,\X_\mathrm{i})$,
$s(T,\rho,\X_\mathrm{i})$ and $\epsilon_\nu(T,\rho,\X_\mathrm{i})$, are the specific nuclear energy generation rate\footnote{{Nuclear reaction rates must account for the screening of the nuclear charge by the electron gas} \citep[e.g.][]{Salpeter54}.}, specific entropy and specific neutrino energy production rate (or neutrino luminosity) and are all function of temperature, density, elemental abundance $\X\mathrm{_i}$ at a given position within the star,  and where specific means that the quantities are evaluated per unit mass, so that they are all expressed in units of [erg~g$^{-1}$~s$^{-1}$]\footnote{Terms not included here are e.g. kinetic energy rates associated with mass loss or mass accretion rates, heating and/or tidal effects by binary companions,  energy production rates by exotic dark matter interactions, etc.}. 

\subsection{The chemical abundances}

\begin{BoxTypeA}[chap3.5:box1]{{The mean molecular weight}}

The number of particles per unit volume, ${N=\NUM/V}$, is often expressed by means of the mean molecular weight $\mu$, which is the average mass per particle in AMU \citep[e.g.][]{Kippenhahn2013}.
We may derive the mean molecular weight of a mixture of elements by considering that
\begin{equation}
    {N}=\mathrm{\frac{\rho}{\mu  \AMU}=\su{i}{\textit{N}_i\times \textit{n}_i}=\frac{\rho}{\AMU} \su{i}{\frac{\X_i}{\Ai}\times \textit{n}_i}} ~,
\end{equation}
where $N\mathrm{_i}$ is given by \eq{\ref{eq:nxyzi}} and  $n\mathrm{_i}$ denotes the number of individual particles provided by the ions ${i}$ (i.e. the nucleus plus the free electrons in case of ionization).
Thus,
\begin{equation}
\frac{1}{\mu}=\mathrm{\su{i}{\frac{\X_i}{\Ai}\times \textit{n}_i}} ~.
\end{equation}
Making use of the mean molecular weight, the pressure of a perfect gas  becomes
\begin{equation}
    {P}={\kb~\frac{\rho T}{\mu~\AMU}} ~.
\end{equation}
For a fully ionized medium
\begin{equation}
\frac{1}{\mu}\simeq~2{\frac{\X}{A_\mathrm{H}}+3\frac{\Y}{A_\mathrm{He}}+ \su{i}{\frac{\Z_\mathrm{i}}{\Ai}\times (\mathrm{\Zch_i}+1)}} ~,
\end{equation}
where $\mathrm{\Zch_i}$ and $\Z_\mathrm{i}$ are the atomic number (charge) and the abundance by mass of the ${i}$ element other than H and He, respectively.
A good approximation is to set
$A\mathrm{_H}$=1, $A\mathrm{_{He}}$=4 and $\mathrm{\Zch_i}+1\simeq~\mathrm{\Ai}/2$ so that
\begin{equation}
    \frac{1}{\mu}\simeq~\mathrm{2\X+\frac{3}{4}\Y+
\su{i}{\frac{\Z_i}{\Ai} \times \frac{\Ai}{2}}
=2\X+\frac{3}{4}\Y+ \frac{\su{i}{\Z_i}}{2}=2\X+\frac{3}{4}\Y+ \frac{\Z}{2}} ~.
\end{equation}
If we are interested in the electron pressure and we only count electrons then,
for full ionization of the same mixture, we have \begin{equation}
\frac{1}{\mu_{\mathrm{e}}}\simeq~\mathrm{\X+\frac{2}{4}\Y+\su{i} 
{\frac{\Z_i}{\Ai} \times \frac{\Ai}{2}}
=\X+\frac{1}{2}\Y+ \frac{\su{i} {\Z_i}}{2}=\X+\frac{\Y}{2}+ \frac{\Z}{2}=\frac{1+\X}{2}} ~,
\label{eq:mue}
\end{equation}
and, for ions \begin{equation}
\frac{1}{\mu_\text{ions}}\simeq~\mathrm{\X+\frac{\Y}{4}+\su{i}{\frac{\Z_i}{\Ai}}} ~.
\end{equation}

\end{BoxTypeA}

In stellar astrophysics the capital letters $\X$, $\Y$ and $\Z$ are used to indicate the fractional abundance by mass of H, He and all remaining elements, called \say{metals}, at a given position ${r}$ 
within the star, respectively. Because by definition $\X = \X_\mathrm{H}=\rho_\mathrm{H}/\rho$ and  similar definitions hold for He and metals, 
\begin{equation}
    \X+\Y+\Z=\frac{\rho_\mathrm{H}+\rho_\mathrm{He}+\rho_\mathrm{M}}{\rho}=1~.
\label{eq:xyz}
\end{equation}

If we follow the abundances of specific elements in detail then
we need to write $\sum{\X \mathrm{_i}}=1$,
and now $\X \mathrm{_i}$ refers to the $\mathrm{i^{th}}$ element in our list.
The equation of state requires the knowledge of the number of atoms of the element  ${i}$  per unit volume, $N\mathrm{_i}$. This can be derived by
dividing  the density of the element $i$, $\rho_\mathrm{i}$, by its mass in grams
\begin{equation}
    N\mathrm{_i}=\frac{\rho_\mathrm{i}}{\Ai \AMU}=
\rho\frac{\X_\mathrm{i}}{\Ai \AMU} ~,
\label{eq:nxyzi}
\end{equation}
where $\Ai$ is the (atomic) mass of element $i$ expressed in Atomic Mass Unit [AMU], $\AMU$ \footnote{$\AMU$=$\mathrm{1/N_A}$, where $\mathrm{N_A}$ is the Avogadro Number.}.
The quantity $\X \mathrm{_i}/\Ai$  is often indicated with \Y$\mathrm{_i}$ which, by definition, indicates  the number of {moles} per gram of the element ${i}$.\\

\subsection{Solving the system of equations of the stellar structure}
The set of the four partial differential equations
(\ref{eq:mass_conservation},
\ref{eq:hydro},
\ref{eq:rtransportross},
\ref{eq:luminosity})
involves four dependent functions of radius and time, ${m(r,t)}$, ${P(r,t)}$,  ${T(r,t)}$ and ${\lum(r,t)}$. 
If the quantities ${\rho(P,T,\X_\mathrm{i}), ~ 
\epsim}$, and ${K_\mathrm{R}(T,\rho, \X\mathrm{_i})}$ can be calculated at each position ${r}$ inside the star, the system of four partial differential equations
(\ref{eq:mass_conservation},
\ref{eq:hydro},
\ref{eq:rtransportross},
\ref{eq:luminosity})
can be integrated to give the internal structure of the star at a given time, 
once four suitable boundary conditions are specified. 
Concerning the time derivative, we note that it appears explicitly in \eq{\ref{eq:hydro_acc}}, when the condition of hydrostatic equilibrium is violated; in the time dependence of the entropy term in \eq{\ref{eq:epsim}}; in the equations governing the variation of the chemical abundances due to nuclear reactions and mixing.
Concerning hydrostatic equilibrium we note that only in the final phases must the dynamical timescales be accounted for, or when particular effects are investigated, such as pulsations in variable stars.
In all the other main phases of stellar evolution, including those of rapid contraction or expansion, the dominant timescales are much longer than the dynamical ones and the
assumption of hydrostatic equilibrium is an excellent approximation. For the above reason, the entropy variation due to  contraction or expansion, \eq{\ref{eq:epsim}}, can be discretized as
\begin{equation}
\pdv{s(r,t)}{t}=\mathrm{\frac{s(r,t_j)-s(r,t_{j-1})}{t_j-t_{j-1}}}~,
\label{eq:epsimdisc}
\end{equation}
with $t\mathrm{_j}$, and $t\mathrm{_{j-1}}$  being the current and the previous time step, and with all quantities at $t\mathrm{_{j-1}}$ being already known from the solution of the previous model.

Instead, the variation of the chemical abundance (number of atoms ${i}$ per unit volume, $N\mathrm{_i}$) due to nuclear reactions and/or  mixing, must be explicitly followed with suitable time-dependent equations such as
\begin{equation}
    \pdv{N_\mathrm{i}}{t}=\mathrm{\frac{\rho}{\Ai \AMU}\pdv{\X\mathrm{_i}}{\textit{t}}= -\su{j} (1+\delta_{ij})\rat_{ij}+\su{k}\su{l} \rat_{kl,i}}+\text{mixing terms}~.\label{eq:abundance_equation}
\end{equation}
In \eq{\ref{eq:abundance_equation}}, the term $\mathrm{\rat_{ij}}$ indicates the nuclear reactions between nuclei $j$ and $i$, that destroy the element $i$ \footnote{When $i=j$, 1+$\mathrm{\delta_{ij}}=2$, because in this case two   elements of type ${i}$ are destroyed in each collision.} while,  the term $\mathrm{\rat_{kl,i}}$ refers to all the collisions between nuclei ${k}$ and ${l}$ that produce a new nucleus ${i}$. Here {\sl mixing terms} refers to all other processes, besides nuclear reactions, that may change the chemical composition of 
a given stellar layer. The most common among such processes  are convection, advection, diffusion, thermohaline mixing, and meridional circulation and shear mixing in  rotating stars \citep[see e.g.][and references therein]{Maeder2009}. 
Usually, a homogeneous  chemical composition \citep[e.g.][for the solar composition]{Asplund21}, that is independent of the position within the star, is adopted for the initial model of a stellar evolution track. This is a reasonable assumption since the molecular gas out of which stars form seems well-mixed. 

Opacity tables $\kros{(T,\rho, \X\mathrm{_i})}$ are provided by several authors \citep[e.g.][]{Iglesias1996,Seaton2005,Hui-Bon-Hoa2021, Marigo2024}; recent comprehensive compilations of nuclear reaction rates and resulting  ${\epsilon_{\mathrm{nuc}}(T,\rho,\X\mathrm{_i})}$  are provided  by \cite{Cyburt2010, XU201361};
electron neutrino energy losses,  ${\epsilon_\nu(T,\rho,\X \mathrm{_i})}$, may be found in \cite{Munakata1985,ItohKohyama1983,Haft1994}, for the different processes.

\subsection{Boundary conditions}
Two boundary conditions can be easily derived because, at the center, ${r}=0$,  
${\lum=0}$ and ${m=0}$.
To obtain the other two boundary conditions for $P$ and $T$ we need to proceed in another way.
Using the results of the 
theory of stellar atmospheres \cite[e.g.][]{Mihalas1970} it is possible to show that, for the particular class of atmosphere models that adopt the grey Eddington approximation, the gas temperature depends on the optical depth $\tau$ in the following way:
\begin{equation}
T^4\left(\tau\right)=\frac{3}{4}\teff^4\left(\tau+\frac{2}{3}\right)~.
\end{equation} 
Thus   $\teff$ corresponds to the physical temperature at the stellar radius $R$ where, by definition, $\tau\mathrm{=}\frac{2}{3}$, and
the surface boundary condition for $T$ is obtained by simply substituting
$\teff$ with $T(r=R)$ in  \eq{\ref{eqteff}}. The boundary condition for $P(r=R)$ is obtained from the same atmosphere models, integrating the equation of the hydrostatic equilibrium from $\tau\mathrm{=}0$ to  $\tau\mathrm{=}\frac{2}{3}$ \citep{Kippenhahn2013}.
Effects of adopting different surface boundary conditions are thoroughly discussed in literature \citep[e.g.][]{VandenBerg08, Chen2014}.

Early powerful numerical methods to obtain \say{\textit{time sequences of stellar models describing evolutionary changes}} \citep{Henyey1964}, i.e. stellar evolution tracks,  have been developed since the middle of the last century \citep{Henyey1959,Hofmeister1964}.
A non-comprehensive list of popular codes to investigate 
stellar evolution includes
ATON \citep{Ventura2008}, 
BASTI \citep{salaris_basti22}, 
CESAM \citep{Morel2008}, 
CLES \citep{Scuflaire2008}, 
FRANEC \citep{Chieffi1998}, 
GARSTEC \citep{Weiss2008},  
GENEC \citep{Eggenberger2008}, 
KEPLER \citep{Weaver1978}, %
KYP2009 \citep{KYP2009}, %
MESA \citep{Paxton2011}, 
PARSEC \citep{Bressan2012}, %
STARS \citep{Eggleton1971}, %
YREC \citep{Demarque2008}\footnote{Updated or newer versions of the listed codes may be found in literature as well as other codes not in this short list.}.
\subsection{The Hertzsprung-Russell diagram}
\label{subsecHRD}
Once an evolutionary track is computed, its
time sequence of luminosity and effective temperature
are usually plotted in the Hertzsprung-Russell
diagram (HRD). The HRD has been independently discovered by Ejnar Hertzsprung and Henry Norris Russell \citep[e.g.][]{Russell1914}.
A Colour-Magnitude diagram (CMD) is obtained if, instead of using $L$ and $\teff$, their observational counterparts,  magnitude and  colour, are adopted.
To convert from a HRD to a CMD, tables of bolometric corrections are needed to translate $\teff$, $L$, and surface gravity into magnitudes and colours \citep[e.g.][]{Chen19}. 
The CMD (Figure~\ref{fig:gaiacm}) and the  corresponding HRD (Figure~\ref{fig:hrd}), are among the most useful  diagnostic tools in stellar astrophysics. They inspired and guided astrophysicists in building the theory of stellar structure and evolution.
\begin{figure}[]
\centering
\caption{The \href{https://aliveuniverse.today/images/articoli/2022/IoW20220523_CMDGaiaJohnsonKronCousin.png}{CMD of GAIA Data Release 3 stars}  (Credits: ESA//Gaia//DPAC  - \href{https://creativecommons.org/licenses/by-sa/3.0/igo/}{CC BY-SA 3.0 IGO}. Acknowledgements: M. Bellazzini - Improvement: Marco Di Lorenzo). Stars with Galactic latitude  $\left|\mathrm{b}\right|>$50$^{\circ}$ (Galactic Caps sample) and with relative distance error less than 10\%, were extracted from the DR3 Gaia Synthetic Photometry Catalogue (GSPC). These stars are representative of different stellar populations in our Galaxy, in particular for their different initial metallicity, \Z. Besides the MS of un-evolved (but nevertheless old) stars, many individual branches clearly stand out in this diagram. These indicate other important stellar evolution phases such as, SUB Giants,  RGB stars and  RGB bump stars, HB stars,  AGB stars with  AGB Bump stars and the WD stars. All these features will be discussed below.}\label{fig:gaiacm}
\includegraphics[width=.70\textwidth]{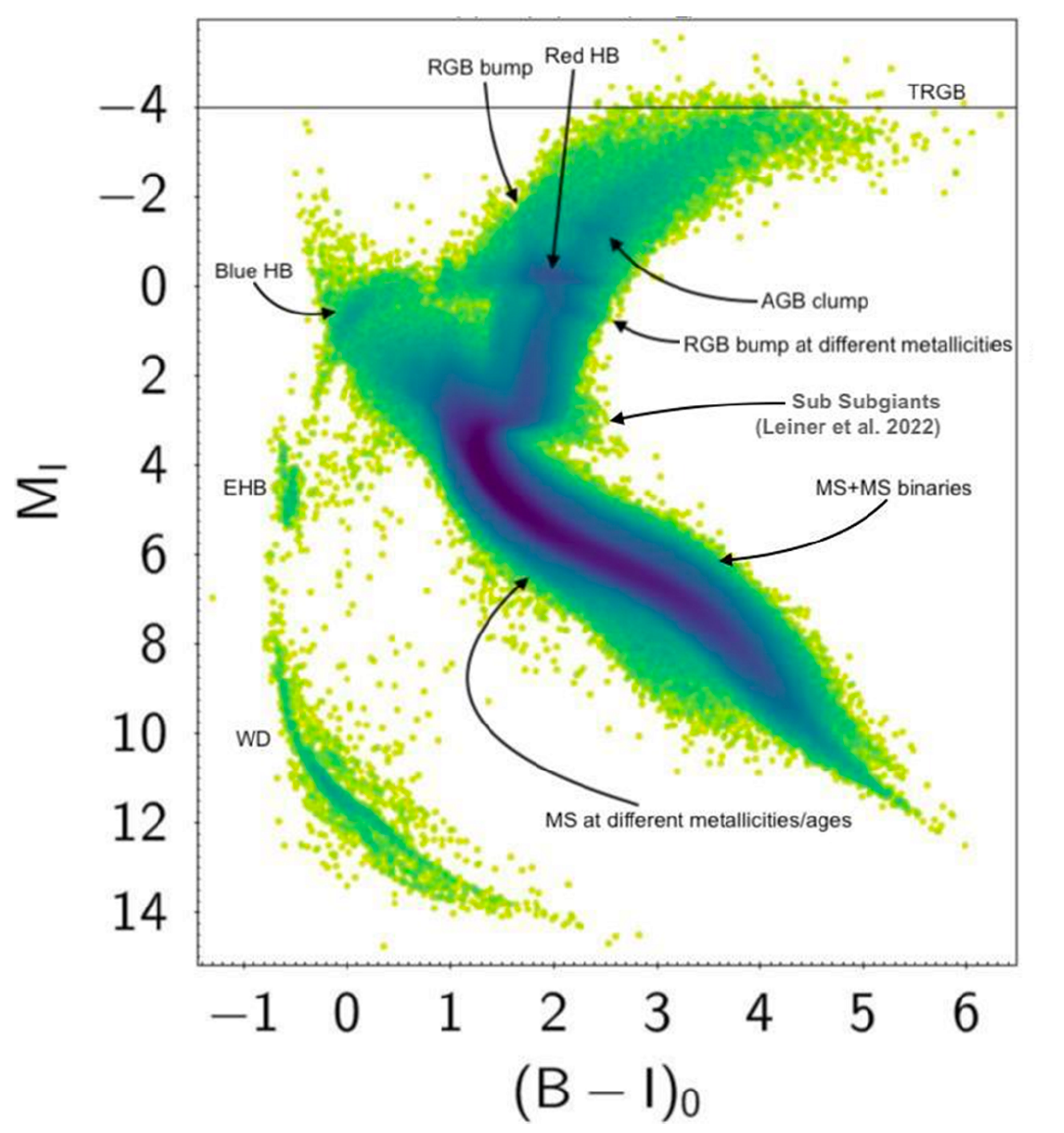}\end{figure}
\begin{figure}[]
\centering
\caption{A Hertzsprung-Russell Diagram of almost solar  metallicity stellar evolutionary tracks at varying initial mass, with models by \cite{Bressan2012,Chen2014}, \citet[][for Post AGB stars]{Millerbertolami} and \citet[][for White Dwarfs]{salaris_basti22}. This diagram is the theoretical foundation of the tools used to decipher the information contained in the CMD of Figure \ref{fig:gaiacm}. Details on the different evolutionary phases are provided in the text.}\label{fig:hrd}
\includegraphics[width=.80\textwidth]{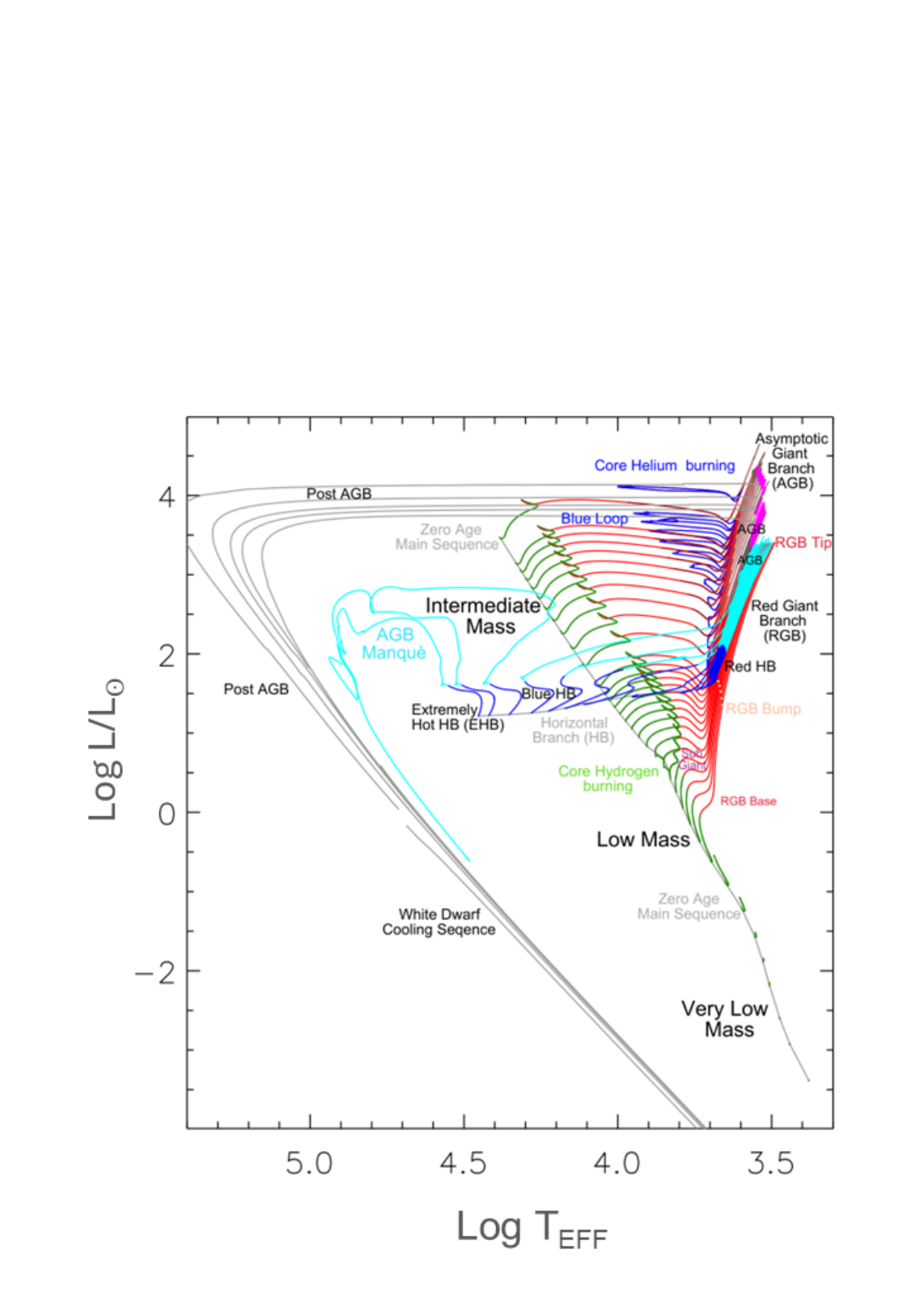}
\end{figure}

\begin{BoxTypeA}[chap4:box1]{Hydrogen burning nuclear reactions}

The main H burning reactions are the proton proton (PP) cycle and the Carbon, Nitrogen, Oxygen (CNO) cycle. The PP cycle is fully active at temperatures above T $\geq$ several 10$^6$~K and it is the typical nuclear energy generation cycle of main sequence stars with mass $\mini\simeq\msun$. For larger masses, when the central temperature becomes larger than about 20$\times$10$^6$~K and for a non-negligible CNO abundance, the CNO  becomes the more efficient cycle. Since the reaction rates of this cycle have a strong temperature dependence, the nuclear energy is produced almost entirely in the very central regions of the stars and it is transported outside by convection. Stars with $\mini\leq 1.1 \msun$ (this value somewhat depending on the chemical composition) have a radiative core with a radiative temperature gradient while, stars with larger $\mini$ have a convective core with an adiabatic temperature gradient.\\

\begin{tabular}{c}
    \multicolumn{1}{c}{\large Proton Proton Cycle} \\
	\hline
    \multicolumn{1}{c}{PPI} \\
    \hline
    \reac{p}{}{p}{^2H}{e^+ +\nu}{} \\
    \reac{p}{}{^2H} {^3He}{\gamma}{} \quad ({\sl Low T})\\
    \reac{He}{3}{^{3}He}{^4He}{2\,p} {}{} ({\sl alt. PPII})\\
	\hline
    \multicolumn{1}{c}{PPII} \\
      \hline
      $^3$He + $^4$He $\rightarrow$ $^7$Be + $\gamma$  \\
      $^7$Be + e$^- \rightarrow$ $^7$Li + $\nu$ $\quad$ ({\sl alt. PPIII})\\
      $^7$Li+p $\rightarrow ^4$He+ $^4$He $\quad$  ({\sl Low T})\\
	\hline
    \multicolumn{1}{c}{PPIII} \\
     \hline
      $^7$Be + p $\rightarrow ^8$B+ $\gamma\quad$ ({\sl Low T})\\
      $^8$B $\rightarrow$ $^8$Be + e$^+$ + $\nu$ \\
      $^8$Be $\rightarrow$ $^4$He + $^4$He \\
      \hline

\end{tabular} \qquad
\renewcommand{\arraystretch}{0.9}
\begin{tabular}[c]{ p{2.5cm} p{0.01mm} p{2.5cm}  }
   \multicolumn{3}{c}{\large  CNO(F) Cycle} \\
	\hline
    \multicolumn{1}{c}{CN}  & & \multicolumn{1}{c}{NF} \\
    \cmidrule(l){1-1} 
    \cmidrule(r){3-3}

    \reac{C}{12}{p}{^{13}N}{\gamma}{}{} & & \reac{O}{17}{p}{^{18}F}{\gamma}{}{}  \\
      $^{13}${N}$\rightarrow$ $^{13}${C} + {e$^+$} + $\nu$ & &$^{18}${F} $\rightarrow$ $^{18}${O} + {e$^+$} + $\nu$ \\
      \reac{C}{13}{p}{^{14}N}{\gamma}{}{} &  & \reac{O}{18}{p}{^{15}N}{^{4}He}{}{} \\
      \reac{N}{14}{p}{^{15}O}{\gamma}{}{} & & \\
     $^{15}${O} $\rightarrow$ $^{15}${N} + {e$^+$} + $\nu$\\
      \reac{N}{15}{p}{^{12}C}{^4He} {} & &\\
	 \cmidrule(l){1-1}  \cmidrule(r){3-3}
    \multicolumn{1}{c}{ON}  & & \multicolumn{1}{c}{OF}\\ 
     \cmidrule(l){1-1} 
    \cmidrule(r){3-3}
      \reac{N}{15}{p}{^{16}O}{\gamma}{}{} & & \reac{O}{18}{p}{^{19}F}{\gamma}{}{} \\
      \reac{O}{16}{p}{^{17}F}{\gamma}{}{} &  & \reac{F}{19}{p}{^{16}O}{^4He}{}{} \\
      $^{17}${F} $\rightarrow$ $^{17}${O} + {e$^+$} + $\nu$\\ 
      \reac{O}{17}{p}{^{14}N}{^{4}He}{}{} & & \\
	\hline

\end{tabular}

In this table, $\mathrm{p}$ is a proton, $\mathrm{\gamma}$ a photon, $\mathrm{e^+}$ a positron, $\mathrm{\nu}$ an electron-neutrino; all other elements are indicated with their symbol and atomic weight.     
{\sl alt. PPII} and {\sl alt. PPIII} indicate the alternative branching channels through which the PP~cycle of reactions may proceed further. An high order PP cycle is favored at a larger  temperature and at a lower H abundance.

{\sl Low T} indicates low temperature reactions that may occur already during the pre-main sequence phase, when the central temperature reaches T=10$^6$~K. They are not energetically important, apart from the first in the list. They are  relevant especially at low metallicity because at low temperatures the star is fully convective and they destroy light elements such as Deuterium, Lithium and Berillium, whose pristine abundances could be important tracers of the Big Bang Nucleo-synthesis environmental conditions.
\end{BoxTypeA}
\section{Protostars and the pre-main sequence phase}\label{pre-main-seq}
Observations of stars in very young clusters, often still embedded in their parent molecular clouds, suggest that stars form in groups containing from a few thousand to many million
objects. The formation involves several complex dynamical and thermal processes
like self-gravity, turbulence, magnetic fields and gas heating and cooling. Altogether, these processes drive the gravitational collapse of the parent molecular clouds and their continuous fragmentation into smaller sub-units, until protostars of very low mass, $\mini \leq 0.01\msun$, are formed in their centres \citep{Binney1987,McKee07}.

In a protostar, the inner regions reach the conditions for  hydrostatic and local thermal equilibrium so that the collapse switches to a contraction. The initial protostar masses are very low, a small fraction of a solar mass, as well as their luminosity which may be less than a thousandth of that of the Sun\footnote{An exception could be that of the very first stars in which, because of the lack of metals in the primordial gas, fragmentation could be less efficient, leading to higher protostellar masses \citep{Klessen23}.}. They are fully convective and they occupy a locus in the  HRD representing the coolest limit for a given star mass and luminosity in hydrostatic equilibrium, the Hayashi limit \citep{Hayashi1961}. The internal regions quickly heat up while matter from outside continuously falls on the surface of the protostar, eventually through a rotating circumstellar disk, increasing its mass (accretion phase). When the central temperature eventually reaches values around 10$^6$ K, some \say{low temperature} nuclear reactions ignite.
   In such objects, the central nuclear burning of Deuterium provides enough energy to slow down the contraction and, since its abundance is continuously refueled by  external accretion and convective mixing, the star has enough time to  increases its mass and its luminosity \citep{Palla93,Kunitomo17}.
Once accretions ends, Deuterium is quickly destroyed and the star contracts and heats up faster. Other nuclear reactions follow, like central Li burning, but the star does not stop its contraction until the main H-burning reaction ignite.
Then the star ends the pre-main sequence phase and reaches the so called Zero-Age Main Sequence (ZAMS), the starting point of evolutionary tracks plotted in Figure \ref{fig:hrd}.
\section{The Main Sequence (MS)}\label{secmainseq}
The position of the observed Main Sequence of stars in the GAIA CMD is indicated by the label in Figure~\ref{fig:gaiacm}. A theoretical ZAMS is shown in Figure~\ref{fig:hrd} for models of about solar metallicity.
In the latter figure, the ZAMS stellar mass increases from bottom ($\mini=0.1\msun$) to top ($\mini=8\msun$). A mass luminosity relation and a $\teff$ luminosity relation are clearly seen in the HRD.

The main H-burning reactions involve the transformation of four H atoms  into a He atom.
The amount of energy released by each entire H fusion process can be easily calculated by considering the corresponding variation of the  {binding energy} of the particle system, before and after the fusion. Its absolute value is obtained as the difference \citep[e.g.][]{Clayton1984}
\begin{equation}
    \Delta~E=(4 m_\mathrm{H}  - m_\mathrm{He}) ~\mathrm{c}^2\simeq 0.007~(4~m_\mathrm{H}~\mathrm{c}^2) \simeq 26.732 ~\mathrm{Mev}~,
    \label{eq:e_h_equation}
\end{equation}
where $m_\mathrm{H}$ and $m_\mathrm{He}$ are the rest masses of H and He, respectively. Equation (\ref{eq:e_h_equation})  shows that the efficiency of the H fusion process is ${{\Delta E}/ {4 m_\mathrm{H} c^2}\sim}$~0.7\%. 
The   {binding energy per nucleon} is computed by dividing this energy by the number of nucleons (protons and neutrons) and it is a well known function of the nucleus mass \citep[e.g.][]{Clayton1984}.
In spite of this very simple scheme, to account for the energy budget provided by nuclear reactions in order to follow the evolution of the star, one needs to explicitly calculate the detailed reaction rates. A non-negligible fraction of the binding energy is taken away by neutrinos when weak interactions are involved in the reactions. This is accounted for by subtracting that energy from the total budget, because the cross  section of the interaction of electron-neutrinos  with the plasma is so small that they freely escape from the star. Moreover one needs to account for the detailed abundance of elements, because of the explicit dependence of the reaction rates from the abundances of different isotopes.

Once such nuclear reactions are ignited in the central regions, the star reaches the so called {\sl secular equilibrium}. Because of the  strong dependence of the reaction rates on the local temperature, if the latter becomes too high, too much energy is generated and the stars expand and cool until they reach  the required temperature at which the pressure gradient  exactly balance the local acceleration of gravity. The opposite happens if the temperature is not sufficiently high. For the same reason, since the reaction rates depend also on the abundance  of the reactants, as H is consumed the reaction rates tend to decrease, but this effect is counteracted
by the contraction induced by self-gravity, that forces the star to reach the right temperature and density to maintain the hydrostatic equilibrium. Secular equilibrium is what makes a star to behave as a perfect nuclear fusion reactor, even for billions of years.

\subsection{The structure of a low mass main sequence star}\label{chap2:lowmassmainseq}
During the Main Sequence (MS) phase, stars convert H into He in their central regions.
Their central temperature and density continue to increase slowly, of just the amount required to maintain the hydrostatic equilibrium in spite of the consumption of the reactants. This causes a secular but limited growth of their luminosity.

Throughout the duration of the MS phase, the inner region of stars with initial mass $\mini \lesssim 1.1~\msun$ is stable against convection. In the outer envelope, the opacity becomes so large that the radiative gradient exceeds  the adiabatic one, and convection sets in.  The MS structure of stars like our Sun is constituted by a  radiative core surrounded by a convective envelope.
In stars with mass  larger than that of the Sun, the radiation flux generated by nuclear reactions is too large to be transported by radiation alone, \eq{\ref{eq:rtransportross}}. Convection is thus the most efficient way to transport energy in their central regions and the new elements produced by nuclear fusion are  continuously  mixed up to the border of the convective core.
At the same time, a mild opacity ensures that their envelopes are radiative. The structure of such stars during the MS phase is characterized by  a convective core surrounded by a radiative envelope. 

While stars inhabit the MS, they obey a certain number of relationships such as the mass  radius-relation, the mass-luminosity relation and the luminosity-effective temperature relation\footnote{The precise form of these relationships depends on the mass range and on the chemical composition.}.  The latter explains why main sequence stars occupy a well defined narrow diagonal  region in the colour-magnitude diagram.
The mass luminosity relation of main sequence stars, which is primarily an observational fact (\textit{\say{...intrinsic brightness and mass are in direct relationship...} \citep{Halm1911,Kuiper1938}})
later explained by the theory, is a fundamental relation that is at the base of the {\sl stellar evolution clock}, namely, the most common methods to measure the age of stellar systems.
The typical timescale of a process involving fuel consumption is
\begin{align}\label{chap5:eq1}
    \tau=\frac{E}{\dot{E}}~,
\end{align}
where ${E}$ is the energy provided by the fuel and $\dot{E} = d{E}/d{t}$ is its rate of consumption. 
For main sequence stars, $d{E}/d{t} = L$  and 
${E=Q f \X M}$ where, ${Q}$ is the amount of energy produced per gram of H that is fused, $\X$ is H abundance and  ${f M}$ is the fraction of the star mass that undergoes H fusion\footnote{${Q}$ is almost constant while ${f}$ depends on the mass of the star.}. Thus, for the nuclear timescale one may write
\begin{align}\label{chap5:eq2}
\mathrm{\tau_{nuc}}=\frac{Q f \X M}{L}~.
\end{align}
Inserting the mass luminosity relation
\begin{align}\label{chap5:eq3}
{L\propto M^\alpha}~,
\end{align}
with $\alpha\sim 3.5$\footnote{The exponent $\alpha $ depends on the stellar mass because of the different efficiency of nuclear reactions and of the different opacities. It decreases toward more massive stars and also toward lower mass stars.}, we get
\begin{align}\label{chap1:htime}
\mathrm{\tau_{nuc}}\propto Q f \X M^{1-\alpha}\propto M^{-2.5}~.
\end{align}
\eq{\ref{chap1:htime}} shows that with increasing mass, the H-burning time decreases strongly.
Since for the Sun $\mathrm{\tau_{nuc}}\simeq$ 10$^{10}$ yr, the largest {\sl nuclear} age of a star of solar composition is \footnote{This is only a rough approximation assuming that the quantities ${f, \alpha, \X}$ do not change. Moreover the H-burning time is taken as the total {\sl nuclear} time of a star because further nuclear reactions are significantly less efficient than H-burning and because these generally occur at a higher luminosity. In astrophysical studies the total time is always computed accounting for all the processes.}
\begin{align}\label{chap1:ntime}
    \mathrm{\tau_{nuc}}(M)\simeq\tau_{\mathrm{nuc}}^\odot ~\left(\frac{M~}{\msun}\right)^{-2.5} ~\mathrm{yr}~. 
\end{align}

\begin{BoxTypeA}[chap5:box1]{The stellar evolution clock}

A stellar generation in a star cluster is assumed to be coeval and chemically homogeneous.
The mass-luminosity relation forces stars of greater mass to exhaust their H fuel faster than stars of lower mass. They thus abandon the main sequence earlier, and also die earlier than less massive stars. Stars in the very upper part of an observed star cluster  main sequence are those that are just abandoning it because they are just exhausting  their central H fuel.  Observationally, this point in the Colour Magnitude diagram is named the turn-off.  The age of stars in the turn-off region, and so that of the star cluster, is equal to their H-burning lifetime, which can be obtained by models once quantities like absolute luminosity, magnitudes, colours and chemical composition are known. One of the goals of stellar evolution theory is to obtain accurate models in such a way that, when compared with observed stars, one can get their properties, in particular their age. This method has been extended to determine ages of composite stellar populations and is nowadays routinely used for determining the ages of entire galaxies up to the largest distances at very high redshift.
\label{chap1:clock}
\end{BoxTypeA}

\section{The Red Giant phase}\label{secrg}
Once H is exhausted in the center, the energy release by nuclear reactions ceases and self-gravity forces the nucleus to contract again. 
By applying the Virial Theorem to low- and intermediate-mass stars \citep[e.g.]{Kippenhahn2013} one may see that  
half of the energy produced by the gravitational contraction ($E_\mathrm{g}$) is spent to increase the internal energy ($E_\mathrm{i}$) to maintain the hydrostatic equilibrium, the other half being radiated away. 
During this phase the fuel is provided by the gravitational potential so that the typical timescale is the gravitational contraction timescale, the so called Kelvin-Helmholtz  timescale \citep{Kippenhahn2013}. It corresponds to the time by which the star will lose its current internal energy with its current luminosity. Since the Virial Theorem, when applied to low- and intermediate-mass stars, states that  
$E_\mathrm{i}$=$\lvert{E_\mathrm{g}}\rvert$/2, we easily get that
\begin{align}\label{chap1:tkh}
\mathrm{\tau_{KH}}=\frac{E_\mathrm{i}~}{L}\simeq\frac{\mathrm{G}M^2}{2 R L}~.
\end{align}
\subsection{The Sub-giant branch (SGB)}\label{chap2:subsec2}
For the Sun, $\mathrm{\tau_{KH}}\simeq$ 3.1 $\times$ 10$^{7}$ yr, which is much shorter than $\mathrm{\tau_{nuc}}$
but much larger than $\mathrm{\tau_{dyn}}$. So, during the contraction phase the stars maintain the hydrostatic equilibrium.
The contraction of the H exhausted core increases the temperature of the surrounding H-rich regions until, very quickly, H burning ignites there. At this point the star is constituted by a nuclear inert  He core surrounded by a H-rich envelope, at the base of which a shell of H-fusion converts H into He. As the time elapses,  the  temperature and density of the He core   increase while the shell feeds new He to the central nucleus.
The higher the core temperature, the larger the H nuclear reaction rates in the shell and the luminosity they provide. This produces an expansion of the H-rich envelope with an initial fast growth of the radius. The surface of the star cools while its size becomes larger and larger. Initially the envelope is radiative and the excess luminosity causes a strong expansion. In the HRD, the star moves at almost constant luminosity toward  the Hayashi limit at lower effective temperatures (the Sub Giant Branch, SGB). As the external envelope cools, the opacity increases, because a larger fraction of the envelope enters the temperature region where H and He partially recombine, where opacity shows a large peak. The convective envelope begins to penetrate into deeper regions than during the previous H-burning phase. Convection is very efficient and it is able to transport all the luminosity from its base to its surface with an adiabatic temperature gradient, without causing a significant expansion. Once convection reaches the central H-poor regions all the luminosity produced by the inner H-burning shell is transported outside without any further expansion. At this point, the path of the star in the HRD turns from horizontal to almost vertical and the excess luminosity generated by the shell is immediately shown at the stellar surface. This turning point defines the base of the Red Giant Branch (RGB) and it is clearly seen in the CMD of well populated star clusters. For the solar model this happens when the convective envelope reaches the inner 50\% of the mass. Needless to say that the effective temperature of this point in the HR diagram depends critically on the ML parameter that determines the efficiency of the convective energy transport. It is interesting to note that with the MLT calibration obtained from the Solar model, the theory is also able to reproduce the observed location of the base of the RGB in the observed CMD of many star clusters\footnote{However, there is increasing evidence, also from hydrodynamic models of stellar convection, that the MLT parameter may slightly depend on the stellar parameters \citep{Trampedach2014}.}.
\subsection{The Red Giant Branch (RGB)}\label{chap6:sec2}
At the beginning of the SGB phase, the He core mass of a solar mass star is only 0.019~$\msun$ with a central density of 1000 g~cm$^{-1}$ and a central temperature 20 million K.
At the end of the SGB phase, the He core mass reaches 0.130~$\msun$, with a central density of 30000 g~cm$^{-1}$ and a central temperature of 26~million~K. However the temperature needed for the ignition of the next nuclear fuel, He, under normal conditions is of about 100 million K.

\begin{BoxTypeA}[chap6.2:box1]{Degeneracy of stellar matter}
\section*{Electron Degeneracy in Stars}
\label{eosNRCD}
In the limiting case of a totally degenerate gas of electrons the equation of state for the non relativistic limit is \citep{Kippenhahn2013}
\begin{align}\label{NRCD}
    {P=\frac{8\pi}{14\mathrm{h}^3m_\mathrm{e}}\left(\frac{3\mathrm{h}^3}{8\pi \AMU}\right)^{5/3}(\rho/\mu_\mathrm{e})^{5/3}\approx 9.99{\times}10^{12}(\rho/\mu_\mathrm{e})^{5/3}}~,
\end{align}
where one should use \eq{\ref{eq:mue}} for $\mu_\mathrm{e}$. 
At increasing density, electrons will become relativistic. For a totally degenerate gas  of relativistic  electrons the pressure is \citep{Kippenhahn2013}
\begin{align}\label{RCD}
    {P=\frac{\mathrm{h c}}{8 \AMU} \left( \frac{3}{\pi \AMU} \right)^{1/3}(\rho/\mu_\mathrm{e})^{4/3}\approx 1.243{\times}10^{15}(\rho/\mu_\mathrm{e})^{4/3}}~.
\end{align}

\subsection*{The mass radius relation of degenerate stars and the Chandrasekhar mass limit}
Considering only electron pressure and combining \eq{\ref{NRCD}} for a completely degenerate non relativistic configuration  with the hydrostatic equilibrium condition \eq{\ref{eq:hydro_inmass}}, we may obtain a relation between the total mass and surface radius
\begin{align}
    9.99{\times}10^{12}(\rho/\mu_\mathrm{e})^{5/3}\simeq {0.5\frac{\mathrm{G}M^2}{4\pi R^4}}&\nonumber\\
9.18{\times}10^{11}{\frac{M^{5/3}}{R^5\mu_{\mathrm{e}}^{5/3}}\simeq 0.5\frac{\mathrm{G}M^2}{4\pi R^4}}&\nonumber\\
   {\left(\frac{M~~}{\msun}\right)^\frac{1}{3}\frac{R~~}{\rsun}\left(\frac{\mu_{\mathrm{e}}}{2}\right)^\frac{5}{3}}&=0.0123~.
 \label{degmr}
\end{align}
This is the mass-radius relation for degenerate configurations, like the inner He core of a RGB star. At increasing mass, the radius decreases and the density increases further. At a certain density electrons become relativistic.
In the limiting case of relativistic complete degeneracy, after combining \eq{\ref{RCD}} with the hydrostatic equilibrium condition \eq{\ref{eq:hydro_inmass}},  the dependence on the radius vanishes and we obtain a constant value for the mass
\begin{align}
   \mch \mathrm{~=\left(\frac{hc}{8 \AMU~G}\right)^{3/2}\frac{3}{\pi \AMU}\frac{1}{\mu_{\mathrm{e}}^2}\sim\frac{5.836}{\mu_{\mathrm{e}}^2}\msun}~.\label{chalimit}
\end{align}
This value was obtained for the first time by \cite{Chandrasekhar35} who discovered that there is a limiting configuration above which totally electron degenerate stars can no longer fulfill the hydrostatic equilibrium condition, {possibly} leading to a collapse of the compact object.
In the case of a mixture without H (He core or CO core) ${\mu_{\mathrm{e}}}\simeq2$ and $\mch \simeq 1.459 \msun$. For Iron, $\mathrm{\mu_e}\simeq2.15$ and $\mch \simeq 1.263 \msun$
\end{BoxTypeA}
The plasma is almost fully ionized and free electrons move with an average kinetic energy which is equal to that of the nuclei (i.e. $E\mathrm{_k=3/2}~ \kb T$), following the energy equipartition principle. 
Another consequence of the Virial Theorem is that, in a contracting core the central values of $\rho$ and $T$ follow the relation 
\begin{equation}
    \rho \mathrm{_{c}}\propto T\mathrm{_{c}}^3~.
\label{eq:rhoTcontract} 
\end{equation}
However, as the matter density increases, the electron gas may become more and more degenerate as a consequence of the Pauli exclusion principle for fermions \citep{Pauli1925}.
Indeed, by comparing \eq{\ref{NRCD}} with the EOS of a non-degenerate ideal gas of electrons, $P\mathrm{_e=\rho}~ \kb T/{\mu_e \AMU} $, we see that the limit where the contribution of  degenerate non relativistic gas exceeds that of non-degenerate electron gas occurs when the density, at a given temperature, exceeds the threshold established by the following relation
\begin{equation}
    {\rho_\mathrm{c}}\propto T\mathrm{_c}^\frac{3}{2}~.     
\label{eq:degnodegnor} 
\end{equation}
\eq{\ref{eq:degnodegnor}} runs flatter than \eq{\ref{eq:rhoTcontract}} and the point of their their  intersection depends on the mass of the contracting core.
Low mass stars are, by definition, those stars for which the intersection of \eq{\ref{eq:rhoTcontract}} and \eq{\ref{eq:degnodegnor}} happens before  reaching the temperature for He ignition, implying that their contracting He nuclei become electron degenerate before they are able to ignite He. 
 Because of the exclusion principle, in a fully degenerate gas of electrons  all the momentum states are occupied up to a threshold value, and new electrons that can be confined in the same elemental volume must occupy higher momentum (energy) states. The electron gas becomes more and more difficult to compress and the equation of state need to account for this effect. Protons and neutrons too obey the same principle but the density at which they begin to be affected is much higher than that of electrons. Thus, while for non-degenerate ions the average kinetic energy per particle continues to be a function of the temperature ($3/2~\kb T$), for the degenerate electrons it becomes progressively higher than that, and more and more dependent on the matter density. The velocities of the electrons become progressively higher at higher densities, their mean free path becomes also higher, and electron conduction may become the most efficient energy transport mechanism, producing large isothermal regions in the deep interiors. The gas remains ideal (only particle collisions are considered) but the equipartition of energy between ions and electrons ceases to hold: electrons acquire an average energy that is several times larger than $3/2~ \kb T$. The gravitational energy of the contraction goes mainly into increasing the electron energy and not that of the ions (their temperature) and the equation of state becomes less and less dependent from the temperature.

If, for some reason, the Sun were to instantly lose its H-rich envelope and remain with 
an He core of 0.130~$\msun$, it would evolve in the following way\footnote{A scenario of this kind could be the result of tight binary evolution. The more massive component is a low mass stars that evolves away from the Main Sequence. While in the SGB or RGB, it fills its {Roche Lobe} and it is spoiled of its H-rich envelope that is partially or totally accreted by the secondary star, which is less massive and still on the main sequence.}.  
The contraction would increase its density until it would become totally degenerate; the whole core will eventually become almost isothermal; energy will be continuously lost by radiation driving the object toward a low luminosity cooling phase that could last for an indefinite time. This will be a prototype of a He White Dwarf.
However, single stars that reach the RGB phase in less than a Hubble time, must have a large H-rich envelope \footnote{The oldest stars with initial mass $\mini\leq~0.85\msun$  have a H-burning lifetime $t\mathrm{_H}$ larger than the  Hubble time $t\mathrm{_{Hub}}\simeq~13.7$Gyr, and so they should still populate the MS.}.
The presence of such H-rich envelope completely changes the subsequent evolution. The H-burning shell at the bottom of the envelope continuously feeds fresh non-degenerate He particles to the electron degenerate He core. While the mass of the He core grows, its average density and temperature increase, in particular that of the non-degenerate gas of  He nuclei. The luminosity and radius of the star increases while, in the HR diagram, the stars climb along the so called Red Giant Branch. The driving mechanism of the evolution is the outward (mass) displacement of the H-burning shell, that occurs in a  nuclear timescale. For this reason the Red Giant Branch of low mass stars is well populated, as can be seen in the CMD of many globular clusters (GCs).

During the ascent on the RGB, many physical processes occur that may leave important observable features. Just after the RGB base, the  convective envelope deepens down to about the inner 20\% of the total star mass. After that point, the structure of the star is well described by an inner electron-degenerate He core, surrounded by a very thin H-burning shell on top of which there is an extended H-rich convective envelope that reaches the photosphere.
The chemical composition within the envelope is almost homogeneous and all the differences between regions that were, or were not, altered by nuclear H fusion in the previous phases, are erased. Convection transports the traces of inner material that were processed by CNO nuclear reactions to the surface (First Dredge-Up phase). One such feature is the conversion of almost all C into N and the other one is the $\mathrm{^{12}C/^{13}C}$ isotopic ratio that, from an initial value of about 100, is reduced  by at least a factor 10. Another feature is the larger He content in the internal regions. What we see at the surface is the result of the dilution of the internal nuclear processed material with the more external original composition\footnote{There is also evidence that, besides the RGB dilution, some other mixing mechanism is active, such as thermohaline mixing \citep{Charbonnel07}.}.
The outward speed of the H-burning shell depends on its temperature and density conditions and on its H abundance. In a well populated star cluster, the observed number of stars at any luminosity (luminosity function) along the RGB, together with their effective temperature and  their surface chemical composition, can be compared with what predicted by models. In this way we may test the robustness of the physical ingredients that we are using in the models, including nuclear reaction rates, neutrino emission processes, opacity of stellar matter, equation of state, convection theory, efficiency of mixing and even assumptions in the numerical procedures\footnote{Effects of interactions with some still unknown type dark matter have also been tested using RGB observations \citep{Raffelt1995, Straniero2020}.}.
For example, when the H-burning shell reaches the border left by the maximum penetration of external convection it finds a steep H profile. At this point, the star spends a certain time to readjust its thermal equilibrium to the higher H abundance of the burning shell. The excess time spent results in a clearly detectable excess of the number of stars 
in that particular point along the RGB, that is known as the RGB Bump. 
As the He core mass grows, the luminosity and the radius of RGB stars increase. They may reach a luminosity that is 2500 times larger than that of the Sun and a radius that is comparable to the Earth's orbit.
Since these high luminosities allow them to be observed even in external galaxies, these stars have been and are the subject of countless astrophysical investigations \citep[e.g.][and citing papers]{Freedman10}.\\

\subsection{The Tip of the Red Giant Branch  (TRGB)}\label{chap1:Heflash}
When the He core approaches a mass $\mhe\sim~0.5\msun$, the inner temperature reaches a value  around 10$^8$~K, which is high enough to cause a relevant fusion of pairs of He nuclei into $\mathrm{^8Be}$.
The resulting excited $\mathrm{^{*8}Be}$ nucleus rapidly decays in its  most likely channel $\mathrm{^{*8}Be}\rightarrow~\mathrm{^4He+^4He}$ giving rise to only a trace of $\mathrm{^8Be}$ equilibrium abundance\footnote{This is the reason why the Big Bang nucleosynthesis could not proceed beyond He and produced only traces of Li and Be \citep{Cyburt16}.}.
However at a density of more than 10$^5$g~cm$^{-1}$ and a temperature of $\sim$10$^8$~K the production of C by the $\alpha$ capture reaction $\mathrm{^8Be+^4He}\rightarrow\mathrm{^{12}C}+\gamma$ becomes relevant.
Since this fusion of three He nuclei into $\mathrm{^{12}C}$ happens almost simultaneously,
this process is named the {\sl triple}$-\alpha$ nuclear burning process.

In the case of secular equilibrium, an excess of nuclear energy  production increases the local temperature that immediately increases the pressure, forcing the surrounding matter to expand and cool to restore an equilibrium condition. The ignition of the 3$\alpha$ nuclear reactions in a star climbing the RGB happens inside a strongly electron degenerate core, for which the dependence of the EOS from the temperature is negligible, $\pdvh{\text{ln}P}{\text{ln}T}\simeq 0$ (see also \eq{\ref{NRCD}} and \eq{\ref{RCD}}). The excess temperature does not produce a local expansion. Instead the local temperature increases, strongly increasing the local release of heat (photons). This cycle might be explosive and, since it should ignite near the center of the star\footnote{At the central temperature and density a certain fraction of electron neutrino/antineutrino pairs are produced that are able to cool the more internal  regions and shift the maximum temperature slightly off-center.}, it could easily destroy the entire star. However the ignition region is not totally degenerate and  convection and conduction drive heat toward less dense regions where electron degeneracy is even less.
Thus, in a very short time-scale, the 3$\alpha$ nuclear reactions produce a strong energy release that is able to expand the surrounding layers, decreasing the degeneracy of electrons until further nuclear burning becomes stable. About 5\% of He is transformed into $^{12}\mathrm{C}$ during this so called  {He-Flash}. Correspondingly, the envelope undergoes a strong contraction and, in the HRD, the star moves from the Tip of the RGB (TRGB) to a lower luminosity and a hotter effective temperature, the so called Horizontal Branch (HB).
\begin{BoxTypeA}[chap6.3:box1]{Main Helium-burning nuclear reactions}
The main He-burning reactions in low- and intermediate-mass stars  are summarized below.\\ 
\begin{tabular}{@{}ll@{} @{}}
     \multicolumn{1}{c}{\large Helium-Burning Reactions} \\
	\hline
      $^4$He + {$^{4}$He}+ {$^{4}$He} $\rightarrow$ $^{12}${C}{+ $\gamma \quad$} {triple~~$\alpha$} \\
      \reac{C}{12}{^{4}He}{^{16}O}{\gamma}{} \\
      \reac{C}{13}{^{4}He}{^{16}O}{n}{}\\
      \reac{N}{14}{^{4}He}{^{18}F}{\gamma}{} \\
      $^{18}${F} $\rightarrow$ $^{18}${O} + {e$^+$} + $\nu$\\
      \reac{N}{15}{^{4}He}{^{19}F}{\gamma}{} \\
      \reac{O}{16}{^{4}He}{^{20}Ne}{\gamma}{}\\
      \reac{O}{17}{^{4}He}{^{20}Ne}{n}{} \\
      \reac{O}{18}{^{4}He}{^{22}Ne}{\gamma}{}\\
	\hline
      \end{tabular}
\\
Reactions with elements of higher atomic number generally require higher temperatures and, 
in low mass stars, the rates for reactions that go beyond $^{16}$O are negligible. Since the $\alpha$ capture reactions have a strong temperature dependence, the nuclear energy during the main He-burning phase is produced almost entirely in the very central regions and it is transported outside by convection. During central He-burning all stars have a convective core with an adiabatic temperature gradient. 
\end{BoxTypeA}

\begin{BoxTypeA}[chap6.3:box2]{The Helium Flash }

The He-flash is a characteristics of all stars with initial mass between 0.5$\msun$ and about $\mhef\simeq~2.0\msun$. The precise value of $\mhef$ depends on other physical stellar parameters like chemical composition and efficiency of mixing. In this mass range, electrons in the central regions become strongly degenerate after central H-burning phase. Stars below 0.5$\msun$ may also become degenerate but they will not be able to reach the temperature for He ignition ($100~ \text{million~K}$). 
During the contraction, stars with higher mass evolve at a higher temperature and lower density, so that stars with $\mini\geq\mhef$
are able to quickly contract to the He ignition temperature,  with a density that never reaches the limit for strong electron degeneracy.  This characteristic is adopted to separate low- and intermediate-mass stars.
Electron degeneracy is also able to prevent H ignition in very low mass stars. Indeed stars with mass below 0.08$\msun$ are not able to ignite H after their pre-main sequence phase and become H Dwarfs that, for their lower surface temperature, are named Brown Dwarfs.
\end{BoxTypeA}

 \section{The Horizontal Branch (HB)}\label{secHB}
The HB is the locus of the HRD where 
low mass stars burn He in their center.
Its name is due to the fact that, using the bolometric magnitude\footnote{The bolometric magnitude measures the luminosity in the whole spectral range. Standard magnitudes, instead, measure the flux received in a relatively narrow range of frequencies, loosing the photons that falls outside this range.},  
it appears to be almost horizontal in the middle of the CMD, sometimes extending from the RGB to beyond the MS.
To understand the almost constant
luminosity of the HB  it is enough to consider that 
all the RGB stars need to reach almost the same He core mass $\mhe \sim~0.5\msun$, to ignite the 3$\alpha$ reactions. Since the luminosity on the HB is mainly due to the He-burning reactions, they have almost the same luminosity. HB stars with different initial mass have different envelope mass. The greater the envelope mass, the larger its total opacity, and the redder the corresponding effective temperature in the HRD. Old Open Clusters have turn-off masses  $\mini\sim~1.5\msun$ and, since $\mhe\sim~0.5\msun$, the  
envelope mass of their He-burning stars are quite large. They thus populate the red region of the HB, the Red-HB. For field low mass stars this region is also named the Red Clump \citep{Girardi2016}. 
Turn-off masses of galactic GCs are around 0.85$\msun$ and their envelope mass should be $\sim$~0.35$\msun$, which still produces a too high opacity to allow the star to populate the Blue-HB side. In order to occupy the Blue-HB, seen in many GCs, stars must have envelopes of a significantly lower mass.
The common view is that old evolved  low mass stars lose a significant fraction of their mass ($\sim~0.2\msun$) while ascending the RGB \citep{Reimers75,Reimers77,Fusipecci75,Schroder05}.

While RGB mass loss could explain the presence of different HB morphologies (Blue or Red-HB) in different GCs, either by means of age differences (older GCs have lower turn-off masses and should appear hotter), or by metallicity differences  (more metal-poor GCs have lower envelope opacities and should be hotter), it is really challenging to explain a composite morphology (Blue and Red-HB) in the same GC, as often observed. 
In fact, there is growing evidence that, contrary to what believed in the past, many galactic and also extragalactic GCs are not homogeneous  stellar systems but show  correlated differences in chemical composition, e.g. Oxygen, Sodium, Aluminum, indicating the presence of multiple stellar generations \citep{milone2022multiple}.
Particularly relevant to determine the HB morphology is the initial He content of different generations, which has been found to vary, in the same GC, by $\mathrm{\delta{\Y}}\sim~0.10$ to $\mathrm{\delta{\Y}}\sim~0.15$ with respect the cosmological nucleo-synthesis value \citep{refId0,10.1093/mnras/stt1993}. A large surface He content may give rise to a Blue-HB and even to an Extreme (hot) HB (EHB), even at high metallicity \citep{Bressan1994}. The reason of this large He spread at a very low metallicity is still under debate \citep{milone2022multiple}.
\begin{BoxTypeA}[chap7:box1]{Stellar Mass loss}

 {Mass-loss} is a mechanism by which a star loses mass during its evolution. For the Sun, the current mass loss rate is estimated to be $\mathrm{\dot{M}\sim 3\times10^{-14}\msun~yr^{-1}}$. Since its H-burning lifetime is estimated to be $t\mathrm{_{H_\odot}\sim 10\times{10^9}yr}$, the total mass lost at this rate would be negligible ($\sim 0.0003\msun$). In other stars mass loss may be larger, e.g. old  GCs  stars loose about 20\% of their initial mass while climbing the RGB. Or even dramatic as in high metallicity supergiant stars, that  may lose more than 80\% of their initial mass before dying \citep{Vink2022}. Mass loss is another important but uncertain process in stellar evolution.
\end{BoxTypeA}

\section{Intermediate mass stars}\label{secims}
Stars more massive than $\mini\geq~\mhef$ have a main sequence evolution that is very similar to that of the upper range of low mass stars. This is because, in both cases, H is burnt in a convective core (i.e. through the CNO cycle). The main sequence mass-luminosity relation implies that their typical H-burning lifetimes are shorter
than those of low mass stars, spanning from one billion years to hundreds of millions of years for intermediate mass stars, and reaching a few million years in massive stars. A formal separation between intermediate mass and massive stars is set at $\mini=\magb\sim~8\msun$, due to the different post-main sequence evolution of stars in the two mass ranges.
The value of $\magb$ depends from several parameters, the most important ones being the elemental abundance and the mixing efficiency of internal convection and, eventually, of rotation.
After the main sequence phase, the He core of stars more massive than $\mini\geq~\mhef$  contracts and quickly reaches central temperatures in excess of 100~million~K, at which stage the triple$-\alpha$ process begins. The typical time scale of the contraction phase, the Kelvin-Helmholtz timescale,  is much shorter than the nuclear time scale and, since the observed number of stars in any post main sequence phase is proportional to the duration of the phase, only very populous  clusters show stars in this phase of their CMD. In galactic Open Clusters, for example, it is unlikely to observe stars in this contraction phase.
During this fast contraction, the H-burning shell has no time to significantly increase the mass of the He core, and the luminosity of the Red Giant Branch at central He ignition, is not much higher than that at the end of the MS. These stars have large convective envelopes and their central He-burning ignites and continues burning in the hotter side of their RGB.
In the main central He-burning phase, the total luminosity is also partly provided by the H-burning shell. Depending on the relative intensity of the He and H nuclear energy sources,  the star may either continue and conclude its He-burning phase near the RGB or, after a certain time that depends on other physical parameters\footnote{Chemical composition, mixing efficiency, efficiency of different He nuclear reaction rates and opacity.}, it may quickly move toward the MS, continuing the core He-burning phase as a Blue/Yellow Giant star. The star performs a Blue Loop in the HR and, for this reason,  stars in this phase are named Blue Loop Stars.
During this loop the star may cross the Cepheid Instability Strip. Inside this strip the envelope becomes dynamically unstable and the star may undergo periodic radial oscillations. Given the dynamical nature of this instability
these stars show a well defined Period-Luminosity relation \citep{Leavitt1912} that constitutes a fundamental  milestone for astronomical distance determinations  and, as such, a milestone for our understanding of the geometry of the Universe.

\begin{BoxTypeA}[chap8:box1]{The period luminosity relation of variable stars.}\\
Starting from \eq{\ref{eq:hydro_timescale}}
and eliminating the radius using the definition of the effective temperature,
${L=4 \pi R^2 \sigma \teff^4}$ 
we obtain for the dynamical timescale $\period$
\begin{equation}
    \period^2={\frac{R^3}{f \G M}=\frac{1}{f \G M \teff^6}\left(\frac{L}{4\pi\sigma}\right)^{\frac{3}{2}}}~.
\label{eq:period_timescale}
\end{equation}
\eq{\ref{eq:period_timescale}} shows that variable stars located within a region of dynamical instability in the HRD follow a characteristic relationship between the  period of the instability and the total mass, the surface luminosity and the effective temperature.  One of such regions is the Cepheid instability strip, from the name of the prototype of this class of stars, Delta Cephei. 
Since this strip is quite narrow and almost vertical in the HR diagram, while the  blue loops are  almost horizontal and  with a well defined mass-luminosity relation $L=L(M)$, this relation becomes a period$-$luminosity relation, see \citep[e.g.][]{Ripepi2019}.
\begin{equation}
    \mathrm{log}_{10}~\period=\alpha(M,\teff)+ \beta~\mathrm{log}_{10}~L ~.
\label{eq:period_luminosity}
\end{equation}
Another similar class is that of RR-Lyrae variables, that are located at the crossing of the instability strip with the HB in GCs \citep{Marconi2015}. An overview of the different classes of variable stars across the HRD can be found in \cite{Eyer08}.
\end{BoxTypeA}

\section{Final nuclear phases of low- and intermediate-mass stars}\label{secfinal}
After central He exhaustion, low- and intermediate-mass stars have a very similar fate. Their structure consists of a central core, composed mainly of C and O (CO core), surrounded by an extended He-rich zone (the fraction of the previous He core above the central convective region) and an external H-rich envelope. The CO core contracts and heats up, going toward the ignition of the next nuclear fuel of C, at about $7\times~10^8$ K. Because of heating produced by contraction, He-burning continues in a shell just above the CO core. The luminosity produced by both contraction and  nuclear reactions expands and cools the whole envelope. In the HRD the stars move asymptotically toward the Red Giant Branch with increasing luminosity and decreasing effective temperature. For this reason the following phase is named the Asymptotic Giant Branch.

\subsection{The Asymptotic Giant Branch}
\label{secagb}
\begin{BoxTypeA}[chap9:box1]{Neutrino energy losses.}

Weak interactions theory predicts that, instead of an
electromagnetic interaction emitting a photon ($\gamma$), an electron neutrino-antineutrino pair ($\nu \bar{\nu}$) is created with a very small relative probability, $\prob$, given by
\begin{equation}
\frac{\prob(\nu\bar{\nu})}{\prob(\gamma)}\simeq3\times10^{-18}{\left(\frac{E_\nu}{m\mathrm{_ec}^2}\right)^4}~,
\label{eq:nuanu_prob}
\end{equation}
where $E_\nu$, the neutrino energy, is typically of the order of the average electron energy, $\sim \kb T$ or even larger in degenerate matter \citep[e.g.][]{Maeder2009}. The neutrino cross section under usual stellar conditions is so small,
$\sigma_\nu~\sim~10^{-44}\mathrm{cm}^2~$, that the neutrino's  mean free path is $\ell_\nu\simeq~1/(n_\mathrm{e}\sigma_\nu)\simeq~10^{20}$cm. Since the mean free path of photons in a dense stellar plasma is of the order of a centimeter, under certain conditions the number of electromagnetic interactions is so high that, in spite of its very small relative probability, the $\nu\bar{\nu}$ process becomes a significant sink of energy. The high temperature dependence implied both by the right hand side term in \eq{\ref{eq:nuanu_prob}} and by the energy density of photons (${\propto T^4}$) indicates that neutrino losses strongly increase with the temperature.
Electron neutrino pairs, $\nu\bar{\nu}$,  can be produced instead of  electron scattering interactions, free-free (Bremsstrahlung) interactions, electron-positron pair production and plasmon production \citep{Haft1994}.
\end{BoxTypeA}
As the CO core contracts,  the central temperature and density increase reaching typical values of ${T_\mathrm{c}}\simeq~2 \times 10^8$ K and $\rho_\mathrm{c}\geq~10^5$g~cm$^{-3}$, where neutrino losses become important. The energy released by gravitational contraction is consumed by neutrinos, thus preventing further heating of the core that enters a phase of strong electron degeneracy.
This fate is common to both low mass stars that undergo the He-flash, and to intermediate mass stars
that ignite He in a non degenerate gas, up to an upper mass limit $\magb\simeq8 \msun$. 


During the early stages (Early Asymptotic Giant Branch)  the He shell quickly moves outward approaching the H shell, that is temporarily extinguished by the expansion and cooling of the envelope. Envelope convection deepens and, in more massive AGB stars, penetrates into the underlying He-rich layers, pushing the bottom  of the H-rich envelope toward the He-rich core. He and N produced by previous H-burning phase are thus transported to the surface (second Dredge-up phase). In a time scale that may last about one tenth of the central He-burning phase the He-burning  shell reaches the base of the H-rich envelope, thus providing enough heating to reignite the H shell. At this point, the He shell quenches off and the star enters the Double Shell Phase (H and He), proceeding with a series of intermittent and alternate ignitions of the H and He shells \citep{Iben1983ARA&A}. H-shell burning increases the intermediate quiescent He-rich region; when the latter reaches a certain mass threshold the He-shell ignites via the triple$-\alpha$ process; ignition is abrupt because heat cannot be efficiently transported away (He thermal pulse), not even by convection that engulfs the He-rich intermediate region with the newly generated C. Ultimately, the heat wave reaches the bottom of the H-rich envelope causing a strong expansion and the quenching of both the H and He shells. The external convection may penetrate into the He-rich region giving rise to the Third Dredge-up phase during which further He-rich matter is conveyed to the surface. In some cases the surface is also enriched by $^{12}$C freshly produced during the thermal pulse. 
Once the heat wave is completely absorbed by the external envelope the latter contracts  and the H shell ignites again, lasting for most of the time (interpulse period), until a further thermal pulse occurs after a thousand or tens of thousands of years \citep{Herwig05} \footnote{A series of  III Dredge-Up episodes may significantly increase the surface $^{12}$C abundance in such a way that the number abundance ratio $^{12}$C/$^{16}$O$>$1. Then, the star appears as a C-(spectral) type star, in net contrast to M-type giants that show $^{12}$C/$^{16}$O$<$1. When the ratio $^{12}$C/$^{16}$O$\simeq$1, the star is of  spectral type S.}. The combined effect of the alternate H-and He-shell burning is that the mass of the degenerate CO core, $\mco$, increases  toward $\mch$ which, for a CO-rich composition is of about 1.4$\msun$. However,  observations and current models suggest that AGB stars lose their whole H-rich envelope before  $\mco$ reaches $\mch$. On one side there are  observations   suggesting the presence of  strong matter outflows leaving the stars at high rates, several $10^{-5}~\msun{~\mathrm{yr}^{-1}}$\citep{Goldman17, Homan18}. 
These high mass loss rates may evaporate an envelope of 5~$\msun$ in a time scale of a few $10^5$ yr, which is much shorter than the nuclear timescale needed for the H- and He-burning shells to increase the core mass to the Chandrasekhar limit \citep{Marigo08}. The other key evidence is the so called White Dwarfs initial-final mass relation 
showing that the progenitors of 
observed White Dwarfs (0.55$\msun \lesssim \mathrm{M_{WD}}\lesssim $1.2$\msun$) are low- and intermediate-mass stars with  0.8$\msun\lesssim~\mini\lesssim$7$\msun$  \citep{Cummings18, Marigo20, Addari24}.\\
\subsection{The Post Asymptotic Giant Branch}\label{secpagb}While stars ascend the AGB, their envelope is removed by strong mass loss. Once the envelope  mass is reduced to a few thousandths of a $\msun$, the surface temperature  begins to increase and the star leaves the AGB. The H-burning shell and the weaker but still efficient mass loss are able to consume/remove the residual H envelope and the star exposes its naked internal core. When its surface temperature reaches 30000~K the photosphere begins to emit energetic photons that are able to ionize the circum-stellar gas previously lost. The stars assume the characteristic morphology of {Planetary Nebulae: nebulae that are ionized by a central source}, which are among the most beautiful
extended objects that can be observed in the sky, an example is shown in Figure \ref{fig:nebula}. The life of these objects is relatively short because their circum-stellar gas is quickly swept away by radiation pressure and at the same time, the central star begins its cooling phase toward the White Dwarf stage \citep{Millerbertolami}.

Stars with initial mass in a narrow mass range $\magb\leq\mini\leq\msagb\simeq\magb+2\msun$, are able to ignite the $^{12}$C+$^{12}$C reaction without experiencing strong electron degeneracy in their CO cores. They produce a $^{12}\mathrm{C}$ exhausted Oxygen/Neon-rich core, that  contracts and evolves toward strong degeneracy. These stars experience a similar final evolution toward the AGB phase but, being more massive and luminous, they are named Super-AGB stars \citep{Siess07}.
\begin{figure}[h]
\centering
\caption{The \href{https:}{Ring Nebula}  (Credits: ESA/Webb, NASA, CSA, M. Barlow (UCL), N. Cox (ACRI-ST), R. Wesson (Cardiff University) and STScI \href{https://creativecommons.org/licenses/by-sa/3.0/igo/}{CC BY-SA 3.0 IGO}. 
The Ring Nebula (also known as M57 and NGC 6720) captured by the James Webb Space Telescope (JWST). The planetary nebula is located about 2,500 ly away in the constellation Lyra, and is formed from a dying low-mass star expelling its outer layers. The ring structure and colors display ionized gas which is illuminated by the central star. The post-AGB central star has a mass M $\simeq$ 0.61 - 0.62 $\msun$ and primarily composed of C and O, with a surface  T$_{\mathrm{eff},*}\simeq$ 120,000 K and luminosity L$_{*} \simeq $ 200 $\lsun$ \citep{ODell2007}, indicating the star is now approaching the WD cooling sequence. 
}
\label{fig:nebula}
\includegraphics[width=.70\textwidth]{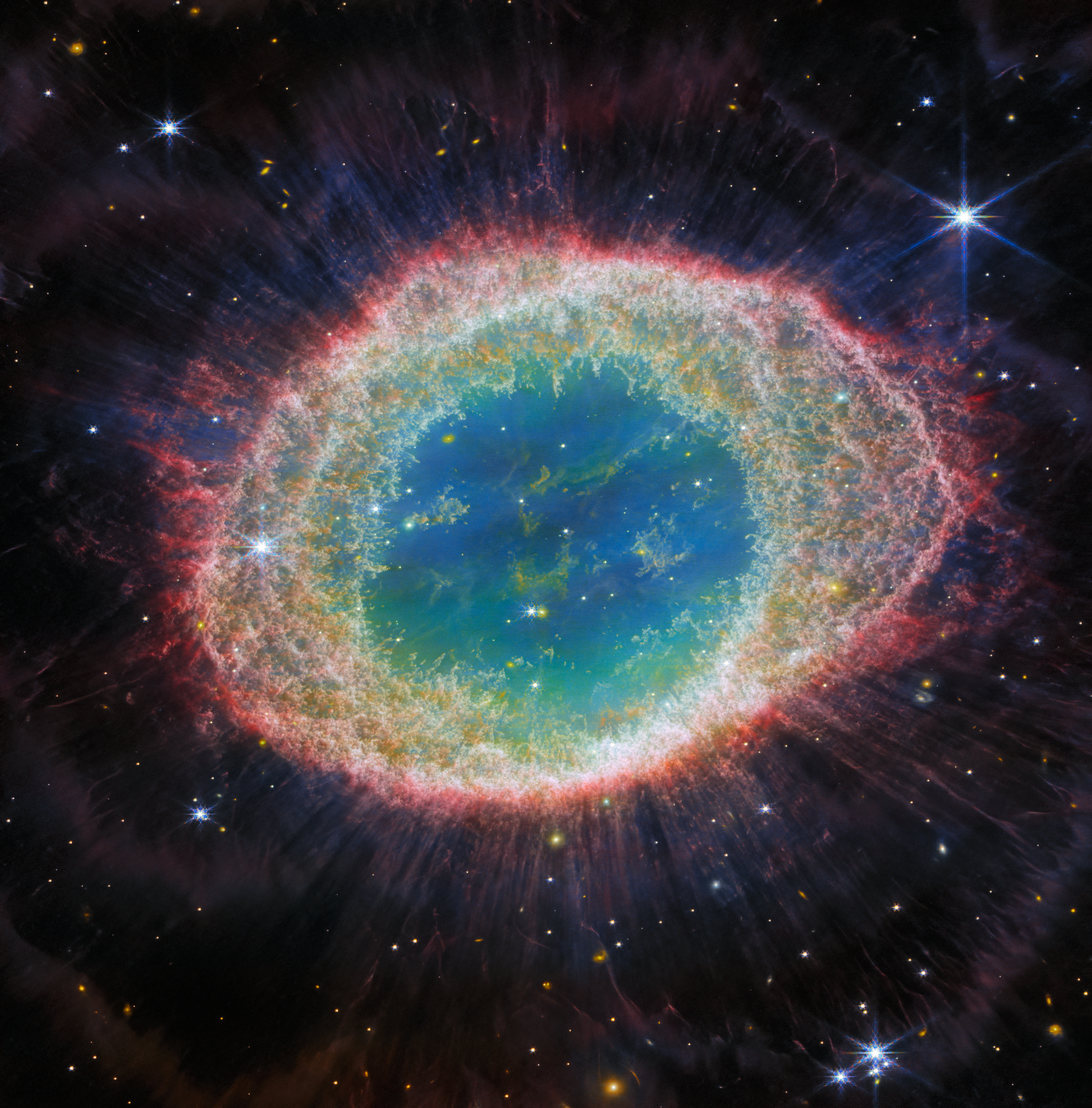}
\end{figure}

\section{White Dwarfs}\label{secwd}
White Dwarfs are thought to be the most common evolved stars because, on one side, all stars with mass $\mini\leq\msagb$  will sooner or later end their nuclear life in that way and, on the other, the lifetime of the WD phase may be even larger than the Hubble time. 
In the HR diagram, the WD sequence is located on the left  (hotter) of the MS, with a luminosity and an effective temperature that decrease almost parallel with that of the MS. However, the higher temperature at any given luminosity suggests that WDs have radii about one hundredth that of a MS star. For a similar mass, this implies a volume that is a millionth, and a density that is a million times that of a MS star with a similar luminosity, respectively. 
This simple argument discloses the power of a well done HR diagram.
Since the early  theoretical interpretations of WDs, it was clear that they must have a very high central density. Indeed, the  fact that a WD must be supported by strongly  degenerate electrons  \citep{Fowler1926} is considered the first application of the Pauli exclusion principle \citep{Kaplan20}.

A completely degenerate configuration is a very good approximation for  WDs, implying that they follow an inverse mass radius relation \eq{\ref{degmr}}, and that there is an asymptotic   upper mass limit, $\mch$, above which they cannot be kept in hydrostatic equilibrium, (see Equation \ref{chalimit}).
For  He- or CO-WDs, $\mu_\mathrm{e}\sim~2/\mathrm{(1+\X)}=2$ and so $\mch\sim~1.46\msun$\footnote{For H-WDs, $\X$=1, $\mu_\mathrm{e}\sim~1$ and  $\mch=5.836\msun$; however H stars with $\mini\geq0.08\msun$ ignite H before degeneracy set in.}.
During the Post-AGB phase, the star quickly crosses the HRD from the coolest to the hottest regions, reaching $\teff$ larger than $\sim$100000 K. 
Then, once nuclear energy generation from the outer very thin H-shell 
quenches off, the star begins to cool again. In this phase, most of the cooling is due to electron-neutrino production by weak interaction from the deep interiors. After a few tens of Myr the interior temperature is no longer high enough for efficient production of neutrinos, and cooling slows down. $\teff$ is now below 30000~K and further luminosity is provided by internal readjustment of the ions gas. 
At increasing density and decreasing temperature, electrostatic interactions become more and more competitive with respect to the thermal energy and force ions, that continue to be non-degenerate, to change from an almost perfect gas phase to liquid and, ultimately, to solid phases. This crystallization process lasts for billions of years. Because of the inverse mass-radius relation, density is higher in more massive objects so that they begin this process earlier and proceed faster, at least in the early stages. 
While the above processes may provide the fuel,  the cooling rate is actually governed by the transport mechanisms across a thin external envelope that, because of its opacity, acts as thermal insulator \citep[e.g.][]{saumon22}. The interior regions are kept almost isothermal by the large mean free path of degenerate electrons\footnote{Their interactions are challenged by unavailability of free momentum states of the outward channel, set by electron degeneracy because all states up to a given energy are already occupied.}. 
Recent cooling tracks for WD stars can be found in \cite{saumon22,salaris_basti22}.
Comparing the observed WD sequences in stellar systems with their theoretical counterparts is a powerful independent  method to estimate their age \citep{Qiu21,tononi2019_wdpopage,Campos16,Bergeron97}.
However the rate at which WD cool down, and the corresponding location in the cooling sequence of the HR diagram, depend strongly on the exact balance between the sources of energy and the efficiency of energy transport at the stellar photosphere, and this may introduce some uncertainties, especially in the timescales of the process  \citep[see e.g.][]{AdamsLaughlin1997}.
Determining the reliability of 
such powerful stellar chronometers is still underway
\citep{Bedard24,salaris_24,Althaus12}.

\section{Conclusions}\label{seconclusions}
This chapter contains a synthetic description of the evolution of low- and intermediate-mass stars. 
The initial stellar mass is the main parameter that determines how stars evolve with time.
It affects the rate of evolution while they are fusing H into He in their hot cores. The larger the initial stellar mass, the larger the luminosity and the shorter the duration of the evolution during the main nuclear burning phases. This property allows a direct determination of the ages of stellar systems. After a long journey on the  main sequence, stars quickly move away when their central H has been totally converted into He. Their central density and temperature quickly increase under the effects of self-gravity. In the mean time their surfaces become very large and cool, while their luminosities increase even by orders of magnitude. In low-mass stars with masses between a fraction to about twice the mass of the Sun, the central density becomes so high that electrons become degenerate before the central temperature reaches the threshold to ignite He-burning reactions, that convert He into C and O. A strongly electron degenerate He core forms and slowly grows in mass because of the surrounding H-burning shell. In this phase, stars climb along the Red Giant Branch. Only when the He core reaches about half a solar mass are the central density and temperature high enough to allow He ignition. At this point the star is at the Tip of the RGB, but given the almost explosive nature of nuclear energy ignition within a strongly degenerate material, the core very quickly expands and the star moves on to the Horizontal Branch in the HRD, where it stays for almost the whole central He-burning phase. Intermediate-mass stars, with initial mass larger than about two solar masses, such as those present in young star clusters, are able to ignite He before meeting the conditions for strong central degeneracy. The RGB of such clusters are thus much less populated and much less developed than that of older clusters that contain evolved low mass stars.
After central He-burning, low- and intermediate-mass stars have a similar structure and follow a common evolution. They are made of a CO core surrounded by a He-rich shell and by an external H-rich envelope. The CO core contracts because of self-gravity, moving toward the ignition of C-burning reactions. However electron neutrino production by weak interactions becomes high enough to consume the energy released by gravitational contraction. The CO core thus becomes degenerate and grows only because of the material processed by the surrounding He- and H-burning shells, that ignite in a series of periodic thermal pulses that characterize the Asymptotic Giant Branch phase.
However, the nuclear growth of the core is challenged by strong stellar winds mainly produced by the formation of dust grains in the cool circum-stellar envelopes of AGB stars. These winds are able to evaporate the whole H-rich envelopes before the CO core reaches the Chandrasekhar mass limit, at which point the star should collapse because not even a strongly electron degenerate core could support its own weight. Instead, a strongly electron degenerate CO core, of mass below $\mch$, surrounded by a very thin He- and eventually H-rich envelope is left by low- and intermediate-mass star, at the end of their nuclear evolution. These stars begin a new life whose duration may be even larger than that of their previous life, as compact objects named White Dwarfs.

Low- and intermediate-mass stars are an essential agent in the cycle of the baryonic matter.  Their evolution  may look less spectacular than that of more massive stars but their number and their persistence are overwhelmingly larger than those of the latter stars, at a level that almost  the whole history of the universe may be recorded and accessed at their surfaces.

A useful database of evolutionary tracks of low- and intermediate-mass stars can be found \href{stev.oapd.inaf.it/PARSEC/index.html}{here}.

\begin{ack}[Acknowledgments]

The authors warmly thank their many close collaborators, Paola Marigo, Leo Girardi, Jing Tang, Yang Chen, Xiaoting Fu, Guglielmo Costa, Thanh Nguyen, Francesco Addari, with whom they continuously discussed open questions in this research field. We also thank Dr. Fabian Schneider and Jan Henneco for their careful reading of the original manuscript and many helpful suggestions. 
\end{ack}

\bibliographystyle{Harvard}
\begin{thebibliography*}{100}
\providecommand{\bibtype}[1]{}
\providecommand{\natexlab}[1]{#1}
{\catcode`\|=0\catcode`\#=12\catcode`\@=11\catcode`\\=12
|immediate|write|@auxout{\expandafter\ifx\csname natexlab\endcsname\relax\gdef\natexlab#1{#1}\fi}}
\renewcommand{\url}[1]{{\tt #1}}
\providecommand{\urlprefix}{URL }
\expandafter\ifx\csname urlstyle\endcsname\relax
  \providecommand{\doi}[1]{doi:\discretionary{}{}{}#1}\else
  \providecommand{\doi}{doi:\discretionary{}{}{}\begingroup \urlstyle{rm}\Url}\fi
\providecommand{\bibinfo}[2]{#2}
\providecommand{\eprint}[2][]{\url{#2}}

\bibtype{Article}%
\bibitem[{Adams} and {Laughlin}(1997)]{AdamsLaughlin1997}
\bibinfo{author}{{Adams} FC} and  \bibinfo{author}{{Laughlin} G} (\bibinfo{year}{1997}), \bibinfo{month}{Apr.}
\bibinfo{title}{{A dying universe: the long-term fate and evolution of astrophysical objects}}.
\bibinfo{journal}{{\em Reviews of Modern Physics}} \bibinfo{volume}{69} (\bibinfo{number}{2}): \bibinfo{pages}{337--372}. \bibinfo{doi}{\doi{10.1103/RevModPhys.69.337}}.
\eprint{astro-ph/9701131}.

\bibtype{Article}%
\bibitem[{Addari} et al.(2024)]{Addari24}
\bibinfo{author}{{Addari} F}, \bibinfo{author}{{Marigo} P}, \bibinfo{author}{{Bressan} A}, \bibinfo{author}{{Costa} G}, \bibinfo{author}{{Shepherd} K} and  \bibinfo{author}{{Volpato} G} (\bibinfo{year}{2024}), \bibinfo{month}{Mar.}
\bibinfo{title}{{The Role of the Third Dredge-up and Mass Loss in Shaping the Initial{\textendash}Final Mass Relation of White Dwarfs}}.
\bibinfo{journal}{{\em \apj}} \bibinfo{volume}{964} (\bibinfo{number}{1}), \bibinfo{eid}{51}. \bibinfo{doi}{\doi{10.3847/1538-4357/ad2067}}.
\eprint{2401.09812}.

\bibtype{Article}%
\bibitem[{Althaus} et al.(2012)]{Althaus12}
\bibinfo{author}{{Althaus} LG}, \bibinfo{author}{{Garc{\'\i}a-Berro} E}, \bibinfo{author}{{Isern} J}, \bibinfo{author}{{C{\'o}rsico} AH} and  \bibinfo{author}{{Miller Bertolami} MM} (\bibinfo{year}{2012}), \bibinfo{month}{Jan.}
\bibinfo{title}{{New phase diagrams for dense carbon-oxygen mixtures and white dwarf evolution}}.
\bibinfo{journal}{{\em \aap}} \bibinfo{volume}{537}, \bibinfo{eid}{A33}. \bibinfo{doi}{\doi{10.1051/0004-6361/201117902}}.
\eprint{1110.5665}.

\bibtype{Article}%
\bibitem[{Asplund} et al.(2021)]{Asplund21}
\bibinfo{author}{{Asplund} M}, \bibinfo{author}{{Amarsi} AM} and  \bibinfo{author}{{Grevesse} N} (\bibinfo{year}{2021}), \bibinfo{month}{Sep.}
\bibinfo{title}{{The chemical make-up of the Sun: A 2020 vision}}.
\bibinfo{journal}{{\em \aap}} \bibinfo{volume}{653}, \bibinfo{eid}{A141}. \bibinfo{doi}{\doi{10.1051/0004-6361/202140445}}.
\eprint{2105.01661}.

\bibtype{Article}%
\bibitem[{Bahng} and {Schwarzschild}(1961)]{Bahng1961}
\bibinfo{author}{{Bahng} J} and  \bibinfo{author}{{Schwarzschild} M} (\bibinfo{year}{1961}), \bibinfo{month}{Sep.}
\bibinfo{title}{{Lifetime of Solar Granules.}}
\bibinfo{journal}{{\em \apj}} \bibinfo{volume}{134}: \bibinfo{pages}{312}. \bibinfo{doi}{\doi{10.1086/147160}}.

\bibtype{Article}%
\bibitem[{Basu} and {Antia}(1997)]{BasuAntia97}
\bibinfo{author}{{Basu} S} and  \bibinfo{author}{{Antia} HM} (\bibinfo{year}{1997}), \bibinfo{month}{May}.
\bibinfo{title}{{Seismic measurement of the depth of the solar convection zone}}.
\bibinfo{journal}{{\em \mnras}} \bibinfo{volume}{287}: \bibinfo{pages}{189--198}.

\bibtype{Article}%
\bibitem[{B{\'e}dard} et al.(2024)]{Bedard24}
\bibinfo{author}{{B{\'e}dard} A}, \bibinfo{author}{{Blouin} S} and  \bibinfo{author}{{Cheng} S} (\bibinfo{year}{2024}), \bibinfo{month}{Mar.}
\bibinfo{title}{{Buoyant crystals halt the cooling of white dwarf stars}}.
\bibinfo{journal}{{\em \nat}} \bibinfo{volume}{627} (\bibinfo{number}{8003}): \bibinfo{pages}{286--288}. \bibinfo{doi}{\doi{10.1038/s41586-024-07102-y}}.

\bibtype{Article}%
\bibitem[{Bergeron} et al.(1997)]{Bergeron97}
\bibinfo{author}{{Bergeron} P}, \bibinfo{author}{{Ruiz} MT} and  \bibinfo{author}{{Leggett} SK} (\bibinfo{year}{1997}), \bibinfo{month}{Jan.}
\bibinfo{title}{{The Chemical Evolution of Cool White Dwarfs and the Age of the Local Galactic Disk}}.
\bibinfo{journal}{{\em \apjs}} \bibinfo{volume}{108} (\bibinfo{number}{1}): \bibinfo{pages}{339--387}. \bibinfo{doi}{\doi{10.1086/312955}}.

\bibtype{Book}%
\bibitem[{Binney} and {Tremaine}(1987)]{Binney1987}
\bibinfo{author}{{Binney} J} and  \bibinfo{author}{{Tremaine} S} (\bibinfo{year}{1987}).
\bibinfo{title}{{Galactic dynamics}}.

\bibtype{Article}%
\bibitem[{B{\"o}hm-Vitense}(1958)]{Bohm-Vitense}
\bibinfo{author}{{B{\"o}hm-Vitense} E} (\bibinfo{year}{1958}), \bibinfo{month}{Jan.}
\bibinfo{title}{{{\"U}ber die Wasserstoffkonvektionszone in Sternen verschiedener Effektivtemperaturen und Leuchtkr{\"a}fte. Mit 5 Textabbildungen}}.
\bibinfo{journal}{{\em \zap}} \bibinfo{volume}{46}: \bibinfo{pages}{108}.

\bibtype{Article}%
\bibitem[{Bressan} et al.(1981)]{Bressan1981}
\bibinfo{author}{{Bressan} AG}, \bibinfo{author}{{Chiosi} C} and  \bibinfo{author}{{Bertelli} G} (\bibinfo{year}{1981}), \bibinfo{month}{Sep.}
\bibinfo{title}{{Mass loss and overshooting in massive stars}}.
\bibinfo{journal}{{\em \aap}} \bibinfo{volume}{102} (\bibinfo{number}{1}): \bibinfo{pages}{25--30}.

\bibtype{Article}%
\bibitem[{Bressan} et al.(1994)]{Bressan1994}
\bibinfo{author}{{Bressan} A}, \bibinfo{author}{{Chiosi} C} and  \bibinfo{author}{{Fagotto} F} (\bibinfo{year}{1994}), \bibinfo{month}{Sep.}
\bibinfo{title}{{Spectrophotometric Evolution of Elliptical Galaxies. I. Ultraviolet Excess and Color-Magnitude-Redshift Relations}}.
\bibinfo{journal}{{\em \apjs}} \bibinfo{volume}{94}: \bibinfo{pages}{63}. \bibinfo{doi}{\doi{10.1086/192073}}.

\bibtype{Article}%
\bibitem[{Bressan} et al.(2012)]{Bressan2012}
\bibinfo{author}{{Bressan} A}, \bibinfo{author}{{Marigo} P}, \bibinfo{author}{{Girardi} L}, \bibinfo{author}{{Salasnich} B}, \bibinfo{author}{{Dal Cero} C}, \bibinfo{author}{{Rubele} S} and  \bibinfo{author}{{Nanni} A} (\bibinfo{year}{2012}), \bibinfo{month}{Nov.}
\bibinfo{title}{{PARSEC: stellar tracks and isochrones with the PAdova and TRieste Stellar Evolution Code}}.
\bibinfo{journal}{{\em \mnras}} \bibinfo{volume}{427} (\bibinfo{number}{1}): \bibinfo{pages}{127--145}. \bibinfo{doi}{\doi{10.1111/j.1365-2966.2012.21948.x}}.
\eprint{1208.4498}.

\bibtype{Article}%
\bibitem[{Buldgen} et al.(2019)]{Buldgen2019}
\bibinfo{author}{{Buldgen} G}, \bibinfo{author}{{Salmon} S} and  \bibinfo{author}{{Noels} A} (\bibinfo{year}{2019}), \bibinfo{month}{Jul.}
\bibinfo{title}{{Progress in global helioseismology: a new light on the solar modelling problem and its implications for solar-like stars}}.
\bibinfo{journal}{{\em Frontiers in Astronomy and Space Sciences}} \bibinfo{volume}{6}, \bibinfo{eid}{42}. \bibinfo{doi}{\doi{10.3389/fspas.2019.00042}}.
\eprint{1906.08213}.

\bibtype{Article}%
\bibitem[{Burrows} and {Vartanyan}(2021)]{Burrows21}
\bibinfo{author}{{Burrows} A} and  \bibinfo{author}{{Vartanyan} D} (\bibinfo{year}{2021}), \bibinfo{month}{Jan.}
\bibinfo{title}{{Core-collapse supernova explosion theory}}.
\bibinfo{journal}{{\em \nat}} \bibinfo{volume}{589} (\bibinfo{number}{7840}): \bibinfo{pages}{29--39}. \bibinfo{doi}{\doi{10.1038/s41586-020-03059-w}}.
\eprint{2009.14157}.

\bibtype{Article}%
\bibitem[{Campos} et al.(2016)]{Campos16}
\bibinfo{author}{{Campos} F}, \bibinfo{author}{{Bergeron} P}, \bibinfo{author}{{Romero} AD}, \bibinfo{author}{{Kepler} SO}, \bibinfo{author}{{Ourique} G}, \bibinfo{author}{{Costa} JES}, \bibinfo{author}{{Bonatto} CJ}, \bibinfo{author}{{Winget} DE}, \bibinfo{author}{{Montgomery} MH}, \bibinfo{author}{{Pacheco} TA} and  \bibinfo{author}{{Bedin} LR} (\bibinfo{year}{2016}), \bibinfo{month}{Mar.}
\bibinfo{title}{{A comparative analysis of the observed white dwarf cooling sequence from globular clusters}}.
\bibinfo{journal}{{\em \mnras}} \bibinfo{volume}{456} (\bibinfo{number}{4}): \bibinfo{pages}{3729--3742}. \bibinfo{doi}{\doi{10.1093/mnras/stv2911}}.
\eprint{1512.03114}.

\bibtype{Article}%
\bibitem[{Cassisi} et al.(2021)]{Cassisi21}
\bibinfo{author}{{Cassisi} S}, \bibinfo{author}{{Potekhin} AY}, \bibinfo{author}{{Salaris} M} and  \bibinfo{author}{{Pietrinferni} A} (\bibinfo{year}{2021}), \bibinfo{month}{Oct.}
\bibinfo{title}{{Electron conduction opacities at the transition between moderate and strong degeneracy: Uncertainties and impacts on stellar models}}.
\bibinfo{journal}{{\em \aap}} \bibinfo{volume}{654}, \bibinfo{eid}{A149}. \bibinfo{doi}{\doi{10.1051/0004-6361/202141425}}.
\eprint{2108.11653}.

\bibtype{Article}%
\bibitem[Chandrasekhar(1935)]{Chandrasekhar35}
\bibinfo{author}{Chandrasekhar S} (\bibinfo{year}{1935}), \bibinfo{month}{01}.
\bibinfo{title}{{The Highly Collapsed Configurations of a Stellar Mass. (Second Paper.)}}.
\bibinfo{journal}{{\em Monthly Notices of the Royal Astronomical Society}} \bibinfo{volume}{95} (\bibinfo{number}{3}): \bibinfo{pages}{207--225}.
ISSN \bibinfo{issn}{0035-8711}. \bibinfo{doi}{\doi{10.1093/mnras/95.3.207}}.
\eprint{https://academic.oup.com/mnras/article-pdf/95/3/207/18326119/mnras95-0207.pdf}, \bibinfo{url}{\url{https://doi.org/10.1093/mnras/95.3.207}}.

\bibtype{Article}%
\bibitem[{Charbonnel, C.} and {Zahn, J.-P.}(2007)]{Charbonnel07}
\bibinfo{author}{{Charbonnel, C.}} and  \bibinfo{author}{{Zahn, J.-P.}} (\bibinfo{year}{2007}).
\bibinfo{title}{Thermohaline mixing: a physical mechanism governing the photospheric composition of low-mass giants}.
\bibinfo{journal}{{\em \aap}} \bibinfo{volume}{467} (\bibinfo{number}{1}): \bibinfo{pages}{L15--L18}. \bibinfo{doi}{\doi{10.1051/0004-6361:20077274}}.
\bibinfo{url}{\url{https://doi.org/10.1051/0004-6361:20077274}}.

\bibtype{Article}%
\bibitem[{Chen} et al.(2014)]{Chen2014}
\bibinfo{author}{{Chen} Y}, \bibinfo{author}{{Girardi} L}, \bibinfo{author}{{Bressan} A}, \bibinfo{author}{{Marigo} P}, \bibinfo{author}{{Barbieri} M} and  \bibinfo{author}{{Kong} X} (\bibinfo{year}{2014}), \bibinfo{month}{Nov.}
\bibinfo{title}{{Improving PARSEC models for very low mass stars}}.
\bibinfo{journal}{{\em \mnras}} \bibinfo{volume}{444} (\bibinfo{number}{3}): \bibinfo{pages}{2525--2543}. \bibinfo{doi}{\doi{10.1093/mnras/stu1605}}.
\eprint{1409.0322}.

\bibtype{Article}%
\bibitem[{Chen} et al.(2019)]{Chen19}
\bibinfo{author}{{Chen} Y}, \bibinfo{author}{{Girardi} L}, \bibinfo{author}{{Fu} X}, \bibinfo{author}{{Bressan} A}, \bibinfo{author}{{Aringer} B}, \bibinfo{author}{{Dal Tio} P}, \bibinfo{author}{{Pastorelli} G}, \bibinfo{author}{{Marigo} P}, \bibinfo{author}{{Costa} G} and  \bibinfo{author}{{Zhang} X} (\bibinfo{year}{2019}), \bibinfo{month}{Dec.}
\bibinfo{title}{{YBC: a stellar bolometric corrections database with variable extinction coefficients. Application to PARSEC isochrones}}.
\bibinfo{journal}{{\em \aap}} \bibinfo{volume}{632}, \bibinfo{eid}{A105}. \bibinfo{doi}{\doi{10.1051/0004-6361/201936612}}.
\eprint{1910.09037}.

\bibtype{Article}%
\bibitem[{Chieffi} et al.(1998)]{Chieffi1998}
\bibinfo{author}{{Chieffi} A}, \bibinfo{author}{{Limongi} M} and  \bibinfo{author}{{Straniero} O} (\bibinfo{year}{1998}), \bibinfo{month}{Aug.}
\bibinfo{title}{{The Evolution of a 25 M$_{{\ensuremath{\odot}}}$ Star from the Main Sequence up to the Onset of the Iron Core Collapse}}.
\bibinfo{journal}{{\em \apj}} \bibinfo{volume}{502} (\bibinfo{number}{2}): \bibinfo{pages}{737--762}. \bibinfo{doi}{\doi{10.1086/305921}}.

\bibtype{Book}%
\bibitem[{Clayton}(1984)]{Clayton1984}
\bibinfo{author}{{Clayton} DD} (\bibinfo{year}{1984}).
\bibinfo{title}{{Principles of stellar evolution and nucleosynthesis.}}

\bibtype{Article}%
\bibitem[{Cummings} et al.(2018)]{Cummings18}
\bibinfo{author}{{Cummings} JD}, \bibinfo{author}{{Kalirai} JS}, \bibinfo{author}{{Tremblay} PE}, \bibinfo{author}{{Ramirez-Ruiz} E} and  \bibinfo{author}{{Choi} J} (\bibinfo{year}{2018}), \bibinfo{month}{Oct.}
\bibinfo{title}{{The White Dwarf Initial-Final Mass Relation for Progenitor Stars from 0.85 to 7.5 M $_{{\ensuremath{\odot}}}$}}.
\bibinfo{journal}{{\em \apj}} \bibinfo{volume}{866} (\bibinfo{number}{1}), \bibinfo{eid}{21}. \bibinfo{doi}{\doi{10.3847/1538-4357/aadfd6}}.
\eprint{1809.01673}.

\bibtype{Article}%
\bibitem[{Cyburt} et al.(2010)]{Cyburt2010}
\bibinfo{author}{{Cyburt} RH}, \bibinfo{author}{{Amthor} AM}, \bibinfo{author}{{Ferguson} R}, \bibinfo{author}{{Meisel} Z}, \bibinfo{author}{{Smith} K}, \bibinfo{author}{{Warren} S}, \bibinfo{author}{{Heger} A}, \bibinfo{author}{{Hoffman} RD}, \bibinfo{author}{{Rauscher} T}, \bibinfo{author}{{Sakharuk} A}, \bibinfo{author}{{Schatz} H}, \bibinfo{author}{{Thielemann} FK} and  \bibinfo{author}{{Wiescher} M} (\bibinfo{year}{2010}), \bibinfo{month}{Jul.}
\bibinfo{title}{{The JINA REACLIB Database: Its Recent Updates and Impact on Type-I X-ray Bursts}}.
\bibinfo{journal}{{\em \apjs}} \bibinfo{volume}{189} (\bibinfo{number}{1}): \bibinfo{pages}{240--252}. \bibinfo{doi}{\doi{10.1088/0067-0049/189/1/240}}.

\bibtype{Article}%
\bibitem[Cyburt et al.(2016)]{Cyburt16}
\bibinfo{author}{Cyburt RH}, \bibinfo{author}{Fields BD}, \bibinfo{author}{Olive KA} and  \bibinfo{author}{Yeh TH} (\bibinfo{year}{2016}), \bibinfo{month}{Feb}.
\bibinfo{title}{Big bang nucleosynthesis: Present status}.
\bibinfo{journal}{{\em Rev. Mod. Phys.}} \bibinfo{volume}{88}: \bibinfo{pages}{015004}. \bibinfo{doi}{\doi{10.1103/RevModPhys.88.015004}}.
\bibinfo{url}{\url{https://link.aps.org/doi/10.1103/RevModPhys.88.015004}}.

\bibtype{Article}%
\bibitem[{Demarque} et al.(2008)]{Demarque2008}
\bibinfo{author}{{Demarque} P}, \bibinfo{author}{{Guenther} DB}, \bibinfo{author}{{Li} LH}, \bibinfo{author}{{Mazumdar} A} and  \bibinfo{author}{{Straka} CW} (\bibinfo{year}{2008}), \bibinfo{month}{Aug.}
\bibinfo{title}{{YREC: the Yale rotating stellar evolution code. Non-rotating version, seismology applications}}.
\bibinfo{journal}{{\em \apss}} \bibinfo{volume}{316} (\bibinfo{number}{1-4}): \bibinfo{pages}{31--41}. \bibinfo{doi}{\doi{10.1007/s10509-007-9698-y}}.
\eprint{0710.4003}.

\bibtype{Article}%
\bibitem[{Eggenberger} et al.(2008)]{Eggenberger2008}
\bibinfo{author}{{Eggenberger} P}, \bibinfo{author}{{Meynet} G}, \bibinfo{author}{{Maeder} A}, \bibinfo{author}{{Hirschi} R}, \bibinfo{author}{{Charbonnel} C}, \bibinfo{author}{{Talon} S} and  \bibinfo{author}{{Ekstr{\"o}m} S} (\bibinfo{year}{2008}), \bibinfo{month}{Aug.}
\bibinfo{title}{{The Geneva stellar evolution code}}.
\bibinfo{journal}{{\em \apss}} \bibinfo{volume}{316} (\bibinfo{number}{1-4}): \bibinfo{pages}{43--54}. \bibinfo{doi}{\doi{10.1007/s10509-007-9511-y}}.

\bibtype{Article}%
\bibitem[{Eggleton}(1971)]{Eggleton1971}
\bibinfo{author}{{Eggleton} PP} (\bibinfo{year}{1971}), \bibinfo{month}{Jan.}
\bibinfo{title}{{The evolution of low mass stars}}.
\bibinfo{journal}{{\em \mnras}} \bibinfo{volume}{151}: \bibinfo{pages}{351}. \bibinfo{doi}{\doi{10.1093/mnras/151.3.351}}.

\bibtype{Inproceedings}%
\bibitem[{Eyer} and {Mowlavi}(2008)]{Eyer08}
\bibinfo{author}{{Eyer} L} and  \bibinfo{author}{{Mowlavi} N} (\bibinfo{year}{2008}), \bibinfo{month}{Oct.}, \bibinfo{title}{{Variable stars across the observational HR diagram}}, \bibinfo{booktitle}{Journal of Physics Conference Series}, \bibinfo{series}{Journal of Physics Conference Series}, \bibinfo{volume}{118}, \bibinfo{publisher}{IOP}, pp. \bibinfo{pages}{012010}, \eprint{0712.3797}.

\bibtype{Article}%
\bibitem[Fowler(1926)]{Fowler1926}
\bibinfo{author}{Fowler RH} (\bibinfo{year}{1926}), \bibinfo{month}{12}.
\bibinfo{title}{{On Dense Matter}}.
\bibinfo{journal}{{\em Monthly Notices of the Royal Astronomical Society}} \bibinfo{volume}{87} (\bibinfo{number}{2}): \bibinfo{pages}{114--122}.
ISSN \bibinfo{issn}{0035-8711}. \bibinfo{doi}{\doi{10.1093/mnras/87.2.114}}.
\eprint{https://academic.oup.com/mnras/article-pdf/87/2/114/3623303/mnras87-0114.pdf}, \bibinfo{url}{\url{https://doi.org/10.1093/mnras/87.2.114}}.

\bibtype{Article}%
\bibitem[{Freedman} and {Madore}(2010)]{Freedman10}
\bibinfo{author}{{Freedman} WL} and  \bibinfo{author}{{Madore} BF} (\bibinfo{year}{2010}), \bibinfo{month}{Sep.}
\bibinfo{title}{{The Hubble Constant}}.
\bibinfo{journal}{{\em \araa}} \bibinfo{volume}{48}: \bibinfo{pages}{673--710}. \bibinfo{doi}{\doi{10.1146/annurev-astro-082708-101829}}.
\eprint{1004.1856}.

\bibtype{Article}%
\bibitem[{Fusi-Pecci} and {Renzini}(1975)]{Fusipecci75}
\bibinfo{author}{{Fusi-Pecci} F} and  \bibinfo{author}{{Renzini} A} (\bibinfo{year}{1975}), \bibinfo{month}{Mar.}
\bibinfo{title}{{On mass loss by stellar wind in population II red giants.}}
\bibinfo{journal}{{\em \aap}} \bibinfo{volume}{39}: \bibinfo{pages}{413--419}.

\bibtype{Article}%
\bibitem[{Girardi}(2016)]{Girardi2016}
\bibinfo{author}{{Girardi} L} (\bibinfo{year}{2016}), \bibinfo{month}{Sep.}
\bibinfo{title}{{Red Clump Stars}}.
\bibinfo{journal}{{\em \araa}} \bibinfo{volume}{54}: \bibinfo{pages}{95--133}. \bibinfo{doi}{\doi{10.1146/annurev-astro-081915-023354}}.

\bibtype{Article}%
\bibitem[{Goldman} et al.(2017)]{Goldman17}
\bibinfo{author}{{Goldman} SR}, \bibinfo{author}{{van Loon} JT}, \bibinfo{author}{{Zijlstra} AA}, \bibinfo{author}{{Green} JA}, \bibinfo{author}{{Wood} PR}, \bibinfo{author}{{Nanni} A}, \bibinfo{author}{{Imai} H}, \bibinfo{author}{{Whitelock} PA}, \bibinfo{author}{{Matsuura} M}, \bibinfo{author}{{Groenewegen} MAT} and  \bibinfo{author}{{G{\'o}mez} JF} (\bibinfo{year}{2017}), \bibinfo{month}{Feb.}
\bibinfo{title}{{The wind speeds, dust content, and mass-loss rates of evolved AGB and RSG stars at varying metallicity}}.
\bibinfo{journal}{{\em \mnras}} \bibinfo{volume}{465} (\bibinfo{number}{1}): \bibinfo{pages}{403--433}. \bibinfo{doi}{\doi{10.1093/mnras/stw2708}}.
\eprint{1610.05761}.

\bibtype{Article}%
\bibitem[{Gray} and {Kaur}(2019)]{Gray19}
\bibinfo{author}{{Gray} DF} and  \bibinfo{author}{{Kaur} T} (\bibinfo{year}{2019}), \bibinfo{month}{Sep.}
\bibinfo{title}{{A Recipe for Finding Stellar Radii, Temperatures, Surface Gravities, Metallicities, and Masses Using Spectral Lines}}.
\bibinfo{journal}{{\em \apj}} \bibinfo{volume}{882} (\bibinfo{number}{2}), \bibinfo{eid}{148}. \bibinfo{doi}{\doi{10.3847/1538-4357/ab2fce}}.

\bibtype{Article}%
\bibitem[{Haft} et al.(1994)]{Haft1994}
\bibinfo{author}{{Haft} M}, \bibinfo{author}{{Raffelt} G} and  \bibinfo{author}{{Weiss} A} (\bibinfo{year}{1994}), \bibinfo{month}{Apr.}
\bibinfo{title}{{Standard and Nonstandard Plasma Neutrino Emission Revisited}}.
\bibinfo{journal}{{\em \apj}} \bibinfo{volume}{425}: \bibinfo{pages}{222}. \bibinfo{doi}{\doi{10.1086/173978}}.
\eprint{astro-ph/9309014}.

\bibtype{Article}%
\bibitem[{Halm}(1911)]{Halm1911}
\bibinfo{author}{{Halm} J} (\bibinfo{year}{1911}), \bibinfo{month}{Jun.}
\bibinfo{title}{{Stars, motion in space, etc. Further considerations relating to the systematic motions of the stars}}.
\bibinfo{journal}{{\em \mnras}} \bibinfo{volume}{71}: \bibinfo{pages}{610--639}. \bibinfo{doi}{\doi{10.1093/mnras/71.8.610}}.

\bibtype{Article}%
\bibitem[{Hayashi}(1961)]{Hayashi1961}
\bibinfo{author}{{Hayashi} C} (\bibinfo{year}{1961}), \bibinfo{month}{Jan.}
\bibinfo{title}{{Stellar evolution in early phases of gravitational contraction.}}
\bibinfo{journal}{{\em \pasj}} \bibinfo{volume}{13}: \bibinfo{pages}{450--452}.

\bibtype{Article}%
\bibitem[{Henyey} et al.(1959)]{Henyey1959}
\bibinfo{author}{{Henyey} LG}, \bibinfo{author}{{Wilets} L}, \bibinfo{author}{{B{\"o}hm} KH}, \bibinfo{author}{{Lelevier} R} and  \bibinfo{author}{{Levee} RD} (\bibinfo{year}{1959}), \bibinfo{month}{May}.
\bibinfo{title}{{A Method for Automatic Computation of Stellar Evolution.}}
\bibinfo{journal}{{\em \apj}} \bibinfo{volume}{129}: \bibinfo{pages}{628}. \bibinfo{doi}{\doi{10.1086/146661}}.

\bibtype{Article}%
\bibitem[{Henyey} et al.(1964)]{Henyey1964}
\bibinfo{author}{{Henyey} LG}, \bibinfo{author}{{Forbes} JE} and  \bibinfo{author}{{Gould} NL} (\bibinfo{year}{1964}), \bibinfo{month}{Jan.}
\bibinfo{title}{{A New Method of Automatic Computation of Stellar Evolution.}}
\bibinfo{journal}{{\em \apj}} \bibinfo{volume}{139}: \bibinfo{pages}{306}. \bibinfo{doi}{\doi{10.1086/147754}}.

\bibtype{Article}%
\bibitem[{Herwig}(2005)]{Herwig05}
\bibinfo{author}{{Herwig} F} (\bibinfo{year}{2005}), \bibinfo{month}{Sep.}
\bibinfo{title}{{Evolution of Asymptotic Giant Branch Stars}}.
\bibinfo{journal}{{\em \araa}} \bibinfo{volume}{43} (\bibinfo{number}{1}): \bibinfo{pages}{435--479}. \bibinfo{doi}{\doi{10.1146/annurev.astro.43.072103.150600}}.

\bibtype{Article}%
\bibitem[{Hofmeister} et al.(1964)]{Hofmeister1964}
\bibinfo{author}{{Hofmeister} E}, \bibinfo{author}{{Kippenhahn} R} and  \bibinfo{author}{{Weigert} A} (\bibinfo{year}{1964}), \bibinfo{month}{Jan.}
\bibinfo{title}{{Sternentwicklung I. Ein Programm zur L{\"o}sung der zeitabh{\"a}ngigen Aufbaugleichungen. Mit 3 Textabbildungen}}.
\bibinfo{journal}{{\em \zap}} \bibinfo{volume}{59}: \bibinfo{pages}{215}.

\bibtype{Article}%
\bibitem[{Homan} et al.(2018)]{Homan18}
\bibinfo{author}{{Homan} W}, \bibinfo{author}{{Richards} A}, \bibinfo{author}{{Decin} L}, \bibinfo{author}{{de Koter} A} and  \bibinfo{author}{{Kervella} P} (\bibinfo{year}{2018}), \bibinfo{month}{Aug.}
\bibinfo{title}{{An unusual face-on spiral in the wind of the M-type AGB star EP Aquarii}}.
\bibinfo{journal}{{\em \aap}} \bibinfo{volume}{616}, \bibinfo{eid}{A34}. \bibinfo{doi}{\doi{10.1051/0004-6361/201832834}}.
\eprint{1804.05684}.

\bibtype{Article}%
\bibitem[{Hui-Bon-Hoa}(2021)]{Hui-Bon-Hoa2021}
\bibinfo{author}{{Hui-Bon-Hoa} A} (\bibinfo{year}{2021}), \bibinfo{month}{Feb.}
\bibinfo{title}{{Stellar models with self-consistent Rosseland opacities. Consequences for stellar structure and evolution}}.
\bibinfo{journal}{{\em \aap}} \bibinfo{volume}{646}, \bibinfo{eid}{L6}. \bibinfo{doi}{\doi{10.1051/0004-6361/202040095}}.

\bibtype{Article}%
\bibitem[{Iben} and {Renzini}(1983)]{Iben1983ARA&A}
\bibinfo{author}{{Iben} I. J} and  \bibinfo{author}{{Renzini} A} (\bibinfo{year}{1983}), \bibinfo{month}{Jan.}
\bibinfo{title}{{Asymptotic giant branch evolution and beyond.}}
\bibinfo{journal}{{\em \araa}} \bibinfo{volume}{21}: \bibinfo{pages}{271--342}. \bibinfo{doi}{\doi{10.1146/annurev.aa.21.090183.001415}}.

\bibtype{Article}%
\bibitem[{Iglesias} and {Rogers}(1996)]{Iglesias1996}
\bibinfo{author}{{Iglesias} CA} and  \bibinfo{author}{{Rogers} FJ} (\bibinfo{year}{1996}), \bibinfo{month}{Jun.}
\bibinfo{title}{{Updated Opal Opacities}}.
\bibinfo{journal}{{\em \apj}} \bibinfo{volume}{464}: \bibinfo{pages}{943}. \bibinfo{doi}{\doi{10.1086/177381}}.

\bibtype{Article}%
\bibitem[{Itoh} and {Kohyama}(1983)]{ItohKohyama1983}
\bibinfo{author}{{Itoh} N} and  \bibinfo{author}{{Kohyama} Y} (\bibinfo{year}{1983}), \bibinfo{month}{Dec.}
\bibinfo{title}{{Neutrino-pair bremsstrahlung in dense stars. I. Liquid metal case.}}
\bibinfo{journal}{{\em \apj}} \bibinfo{volume}{275}: \bibinfo{pages}{858--866}. \bibinfo{doi}{\doi{10.1086/161579}}.

\bibtype{Article}%
\bibitem[Kaplan(2020)]{Kaplan20}
\bibinfo{author}{Kaplan IG} (\bibinfo{year}{2020}).
\bibinfo{title}{The pauli exclusion principle and the problems of its experimental verification}.
\bibinfo{journal}{{\em Symmetry}} \bibinfo{volume}{12} (\bibinfo{number}{2}).
ISSN \bibinfo{issn}{2073-8994}. \bibinfo{doi}{\doi{10.3390/sym12020320}}.
\bibinfo{url}{\url{https://www.mdpi.com/2073-8994/12/2/320}}.

\bibtype{Book}%
\bibitem[{Kippenhahn} et al.(2013)]{Kippenhahn2013}
\bibinfo{author}{{Kippenhahn} R}, \bibinfo{author}{{Weigert} A} and  \bibinfo{author}{{Weiss} A} (\bibinfo{year}{2013}).
\bibinfo{title}{{Stellar Structure and Evolution}}.
\bibinfo{doi}{\doi{10.1007/978-3-642-30304-3}}.

\bibtype{Article}%
\bibitem[Klessen and Glover(2023)]{Klessen23}
\bibinfo{author}{Klessen RS} and  \bibinfo{author}{Glover SC} (\bibinfo{year}{2023}).
\bibinfo{title}{The first stars: Formation, properties, and impact}.
\bibinfo{journal}{{\em Annual Review of Astronomy and Astrophysics}} \bibinfo{volume}{61} (\bibinfo{number}{Volume 61, 2023}): \bibinfo{pages}{65--130}.
ISSN \bibinfo{issn}{1545-4282}. \bibinfo{doi}{\doi{https://doi.org/10.1146/annurev-astro-071221-053453}}.
\bibinfo{url}{\url{https://www.annualreviews.org/content/journals/10.1146/annurev-astro-071221-053453}}, \bibinfo{type}{Journal Article}.

\bibtype{Article}%
\bibitem[{Kovetz} et al.(2009)]{KYP2009}
\bibinfo{author}{{Kovetz} A}, \bibinfo{author}{{Yaron} O} and  \bibinfo{author}{{Prialnik} D} (\bibinfo{year}{2009}), \bibinfo{month}{Jun.}
\bibinfo{title}{{A new, efficient stellar evolution code for calculating complete evolutionary tracks}}.
\bibinfo{journal}{{\em \mnras}} \bibinfo{volume}{395} (\bibinfo{number}{4}): \bibinfo{pages}{1857--1874}. \bibinfo{doi}{\doi{10.1111/j.1365-2966.2009.14670.x}}.
\eprint{0809.4207}.

\bibtype{Article}%
\bibitem[{Kuiper}(1938)]{Kuiper1938}
\bibinfo{author}{{Kuiper} GP} (\bibinfo{year}{1938}), \bibinfo{month}{Nov.}
\bibinfo{title}{{The Empirical Mass-Luminosity Relation.}}
\bibinfo{journal}{{\em \apj}} \bibinfo{volume}{88}: \bibinfo{pages}{472}. \bibinfo{doi}{\doi{10.1086/143999}}.

\bibtype{Article}%
\bibitem[{Kunitomo} et al.(2017)]{Kunitomo17}
\bibinfo{author}{{Kunitomo} M}, \bibinfo{author}{{Guillot} T}, \bibinfo{author}{{Takeuchi} T} and  \bibinfo{author}{{Ida} S} (\bibinfo{year}{2017}), \bibinfo{month}{Mar.}
\bibinfo{title}{{Revisiting the pre-main-sequence evolution of stars. I. Importance of accretion efficiency and deuterium abundance}}.
\bibinfo{journal}{{\em \aap}} \bibinfo{volume}{599}, \bibinfo{eid}{A49}. \bibinfo{doi}{\doi{10.1051/0004-6361/201628260}}.
\eprint{1702.07901}.

\bibtype{Article}%
\bibitem[{Leavitt} and {Pickering}(1912)]{Leavitt1912}
\bibinfo{author}{{Leavitt} HS} and  \bibinfo{author}{{Pickering} EC} (\bibinfo{year}{1912}), \bibinfo{month}{Mar.}
\bibinfo{title}{{Periods of 25 Variable Stars in the Small Magellanic Cloud.}}
\bibinfo{journal}{{\em Harvard College Observatory Circular}} \bibinfo{volume}{173}: \bibinfo{pages}{1--3}.

\bibtype{Article}%
\bibitem[{Low}(2001)]{Low01}
\bibinfo{author}{{Low} BC} (\bibinfo{year}{2001}), \bibinfo{month}{Nov.}
\bibinfo{title}{{Coronal mass ejections, magnetic flux ropes, and solar magnetism}}.
\bibinfo{journal}{{\em \jgr}} \bibinfo{volume}{106} (\bibinfo{number}{A11}): \bibinfo{pages}{25141--25164}. \bibinfo{doi}{\doi{10.1029/2000JA004015}}.

\bibtype{Book}%
\bibitem[{Maeder}(2009)]{Maeder2009}
\bibinfo{author}{{Maeder} A} (\bibinfo{year}{2009}).
\bibinfo{title}{{Physics, Formation and Evolution of Rotating Stars}}.
\bibinfo{doi}{\doi{10.1007/978-3-540-76949-1}}.

\bibtype{Article}%
\bibitem[{Marconi} et al.(2015)]{Marconi2015}
\bibinfo{author}{{Marconi} M}, \bibinfo{author}{{Coppola} G}, \bibinfo{author}{{Bono} G}, \bibinfo{author}{{Braga} V}, \bibinfo{author}{{Pietrinferni} A}, \bibinfo{author}{{Buonanno} R}, \bibinfo{author}{{Castellani} M}, \bibinfo{author}{{Musella} I}, \bibinfo{author}{{Ripepi} V} and  \bibinfo{author}{{Stellingwerf} RF} (\bibinfo{year}{2015}), \bibinfo{month}{Jul.}
\bibinfo{title}{{On a New Theoretical Framework for RR Lyrae Stars. I. The Metallicity Dependence}}.
\bibinfo{journal}{{\em \apj}} \bibinfo{volume}{808} (\bibinfo{number}{1}), \bibinfo{eid}{50}. \bibinfo{doi}{\doi{10.1088/0004-637X/808/1/50}}.
\eprint{1505.02531}.

\bibtype{Article}%
\bibitem[{Marigo} et al.(2008)]{Marigo08}
\bibinfo{author}{{Marigo} P}, \bibinfo{author}{{Girardi} L}, \bibinfo{author}{{Bressan} A}, \bibinfo{author}{{Groenewegen} MAT}, \bibinfo{author}{{Silva} L} and  \bibinfo{author}{{Granato} GL} (\bibinfo{year}{2008}), \bibinfo{month}{May}.
\bibinfo{title}{{Evolution of asymptotic giant branch stars. II. Optical to far-infrared isochrones with improved TP-AGB models}}.
\bibinfo{journal}{{\em \aap}} \bibinfo{volume}{482} (\bibinfo{number}{3}): \bibinfo{pages}{883--905}. \bibinfo{doi}{\doi{10.1051/0004-6361:20078467}}.
\eprint{0711.4922}.

\bibtype{Article}%
\bibitem[{Marigo} et al.(2020)]{Marigo20}
\bibinfo{author}{{Marigo} P}, \bibinfo{author}{{Cummings} JD}, \bibinfo{author}{{Curtis} JL}, \bibinfo{author}{{Kalirai} J}, \bibinfo{author}{{Chen} Y}, \bibinfo{author}{{Tremblay} PE}, \bibinfo{author}{{Ramirez-Ruiz} E}, \bibinfo{author}{{Bergeron} P}, \bibinfo{author}{{Bladh} S}, \bibinfo{author}{{Bressan} A}, \bibinfo{author}{{Girardi} L}, \bibinfo{author}{{Pastorelli} G}, \bibinfo{author}{{Trabucchi} M}, \bibinfo{author}{{Cheng} S}, \bibinfo{author}{{Aringer} B} and  \bibinfo{author}{{Tio} PD} (\bibinfo{year}{2020}), \bibinfo{month}{Jul.}
\bibinfo{title}{{Carbon star formation as seen through the non-monotonic initial-final mass relation}}.
\bibinfo{journal}{{\em Nature Astronomy}} \bibinfo{volume}{4}: \bibinfo{pages}{1102--1110}. \bibinfo{doi}{\doi{10.1038/s41550-020-1132-1}}.
\eprint{2007.04163}.

\bibtype{Article}%
\bibitem[{Marigo} et al.(2024)]{Marigo2024}
\bibinfo{author}{{Marigo} P}, \bibinfo{author}{{Woitke} P}, \bibinfo{author}{{Tognelli} E}, \bibinfo{author}{{Girardi} L}, \bibinfo{author}{{Aringer} B} and  \bibinfo{author}{{Bressan} A} (\bibinfo{year}{2024}), \bibinfo{month}{Jan.}
\bibinfo{title}{{{\AE}SOPUS 2.0: Low-temperature Opacities with Solid Grains}}.
\bibinfo{journal}{{\em \apj}} \bibinfo{volume}{960} (\bibinfo{number}{1}), \bibinfo{eid}{18}. \bibinfo{doi}{\doi{10.3847/1538-4357/ad0898}}.
\eprint{2310.14588}.

\bibtype{Article}%
\bibitem[Marino et al.(2013)]{10.1093/mnras/stt1993}
\bibinfo{author}{Marino AF}, \bibinfo{author}{Milone AP}, \bibinfo{author}{Przybilla N}, \bibinfo{author}{Bergemann M}, \bibinfo{author}{Lind K}, \bibinfo{author}{Asplund M}, \bibinfo{author}{Cassisi S}, \bibinfo{author}{Catelan M}, \bibinfo{author}{Casagrande L}, \bibinfo{author}{Valcarce AAR}, \bibinfo{author}{Bedin LR}, \bibinfo{author}{Cortés C}, \bibinfo{author}{D'Antona F}, \bibinfo{author}{Jerjen H}, \bibinfo{author}{Piotto G}, \bibinfo{author}{Schlesinger K}, \bibinfo{author}{Zoccali M} and  \bibinfo{author}{Angeloni R} (\bibinfo{year}{2013}), \bibinfo{month}{11}.
\bibinfo{title}{{Helium enhanced stars and multiple populations along the horizontal branch of NGC 2808: direct spectroscopic measurements}}.
\bibinfo{journal}{{\em Monthly Notices of the Royal Astronomical Society}} \bibinfo{volume}{437} (\bibinfo{number}{2}): \bibinfo{pages}{1609--1627}.
ISSN \bibinfo{issn}{0035-8711}. \bibinfo{doi}{\doi{10.1093/mnras/stt1993}}.
\eprint{https://academic.oup.com/mnras/article-pdf/437/2/1609/13763705/stt1993.pdf}, \bibinfo{url}{\url{https://doi.org/10.1093/mnras/stt1993}}.

\bibtype{Article}%
\bibitem[{McKee} and {Ostriker}(2007)]{McKee07}
\bibinfo{author}{{McKee} CF} and  \bibinfo{author}{{Ostriker} EC} (\bibinfo{year}{2007}), \bibinfo{month}{Sep.}
\bibinfo{title}{{Theory of Star Formation}}.
\bibinfo{journal}{{\em \araa}} \bibinfo{volume}{45} (\bibinfo{number}{1}): \bibinfo{pages}{565--687}. \bibinfo{doi}{\doi{10.1146/annurev.astro.45.051806.110602}}.
\eprint{0707.3514}.

\bibtype{Book}%
\bibitem[{Mihalas}(1970)]{Mihalas1970}
\bibinfo{author}{{Mihalas} D} (\bibinfo{year}{1970}).
\bibinfo{title}{{Stellar atmospheres}}.

\bibtype{Article}%
\bibitem[{Miller Bertolami, Marcelo Miguel}(2016)]{Millerbertolami}
\bibinfo{author}{{Miller Bertolami, Marcelo Miguel}} (\bibinfo{year}{2016}).
\bibinfo{title}{New models for the evolution of post-asymptotic giant branch stars and central stars of planetary nebulae}.
\bibinfo{journal}{{\em \aap}} \bibinfo{volume}{588}: \bibinfo{pages}{A25}. \bibinfo{doi}{\doi{10.1051/0004-6361/201526577}}.
\bibinfo{url}{\url{https://doi.org/10.1051/0004-6361/201526577}}.

\bibtype{Misc}%
\bibitem[Milone and Marino(2022)]{milone2022multiple}
\bibinfo{author}{Milone AP} and  \bibinfo{author}{Marino AF} (\bibinfo{year}{2022}).
\bibinfo{title}{Multiple populations in star clusters}.
\eprint{2206.10564}.

\bibtype{Article}%
\bibitem[{Morel} and {Lebreton}(2008)]{Morel2008}
\bibinfo{author}{{Morel} P} and  \bibinfo{author}{{Lebreton} Y} (\bibinfo{year}{2008}), \bibinfo{month}{Aug.}
\bibinfo{title}{{CESAM: a free code for stellar evolution calculations}}.
\bibinfo{journal}{{\em \apss}} \bibinfo{volume}{316} (\bibinfo{number}{1-4}): \bibinfo{pages}{61--73}. \bibinfo{doi}{\doi{10.1007/s10509-007-9663-9}}.
\eprint{0801.2019}.

\bibtype{Article}%
\bibitem[{Munakata} et al.(1985)]{Munakata1985}
\bibinfo{author}{{Munakata} H}, \bibinfo{author}{{Kohyama} Y} and  \bibinfo{author}{{Itoh} N} (\bibinfo{year}{1985}), \bibinfo{month}{Sep.}
\bibinfo{title}{{Neutrino Energy Loss in Stellar Interiors}}.
\bibinfo{journal}{{\em \apj}} \bibinfo{volume}{296}: \bibinfo{pages}{197}. \bibinfo{doi}{\doi{10.1086/163436}}.

\bibtype{Article}%
\bibitem[{O'Dell} et al.(2007)]{ODell2007}
\bibinfo{author}{{O'Dell} CR}, \bibinfo{author}{{Sabbadin} F} and  \bibinfo{author}{{Henney} WJ} (\bibinfo{year}{2007}), \bibinfo{month}{Oct.}
\bibinfo{title}{{The Three-Dimensional Ionization Structure and Evolution of NGC 6720, The Ring Nebula}}.
\bibinfo{journal}{{\em \aj}} \bibinfo{volume}{134} (\bibinfo{number}{4}): \bibinfo{pages}{1679--1692}. \bibinfo{doi}{\doi{10.1086/521823}}.

\bibtype{Article}%
\bibitem[{Palla} and {Stahler}(1993)]{Palla93}
\bibinfo{author}{{Palla} F} and  \bibinfo{author}{{Stahler} SW} (\bibinfo{year}{1993}), \bibinfo{month}{Nov.}
\bibinfo{title}{{The Pre-Main-Sequence Evolution of Intermediate-Mass Stars}}.
\bibinfo{journal}{{\em \apj}} \bibinfo{volume}{418}: \bibinfo{pages}{414}. \bibinfo{doi}{\doi{10.1086/173402}}.

\bibtype{Article}%
\bibitem[{Pasquini, L.} et al.(2011)]{refId0}
\bibinfo{author}{{Pasquini, L.}}, \bibinfo{author}{{Mauas, P.}}, \bibinfo{author}{{Käufl, H. U.}} and  \bibinfo{author}{{Cacciari, C.}} (\bibinfo{year}{2011}).
\bibinfo{title}{Measuring helium abundance difference in giants of ngc 2808}.
\bibinfo{journal}{{\em \aap}} \bibinfo{volume}{531}: \bibinfo{pages}{A35}. \bibinfo{doi}{\doi{10.1051/0004-6361/201116592}}.
\bibinfo{url}{\url{https://doi.org/10.1051/0004-6361/201116592}}.

\bibtype{Article}%
\bibitem[{Pauli}(1925)]{Pauli1925}
\bibinfo{author}{{Pauli} W} (\bibinfo{year}{1925}), \bibinfo{month}{Feb.}
\bibinfo{title}{{{\"U}ber den Zusammenhang des Abschlusses der Elektronengruppen im Atom mit der Komplexstruktur der Spektren}}.
\bibinfo{journal}{{\em Zeitschrift fur Physik}} \bibinfo{volume}{31} (\bibinfo{number}{1}): \bibinfo{pages}{765--783}. \bibinfo{doi}{\doi{10.1007/BF02980631}}.

\bibtype{Article}%
\bibitem[{Paxton} et al.(2011)]{Paxton2011}
\bibinfo{author}{{Paxton} B}, \bibinfo{author}{{Bildsten} L}, \bibinfo{author}{{Dotter} A}, \bibinfo{author}{{Herwig} F}, \bibinfo{author}{{Lesaffre} P} and  \bibinfo{author}{{Timmes} F} (\bibinfo{year}{2011}), \bibinfo{month}{Jan.}
\bibinfo{title}{{Modules for Experiments in Stellar Astrophysics (MESA)}}.
\bibinfo{journal}{{\em \apjs}} \bibinfo{volume}{192} (\bibinfo{number}{1}), \bibinfo{eid}{3}. \bibinfo{doi}{\doi{10.1088/0067-0049/192/1/3}}.
\eprint{1009.1622}.

\bibtype{Article}%
\bibitem[{Pr{\v{s}}a} et al.(2016)]{prsa16}
\bibinfo{author}{{Pr{\v{s}}a} A}, \bibinfo{author}{{Harmanec} P}, \bibinfo{author}{{Torres} G}, \bibinfo{author}{{Mamajek} E}, \bibinfo{author}{{Asplund} M}, \bibinfo{author}{{Capitaine} N}, \bibinfo{author}{{Christensen-Dalsgaard} J}, \bibinfo{author}{{Depagne} {\'E}}, \bibinfo{author}{{Haberreiter} M}, \bibinfo{author}{{Hekker} S}, \bibinfo{author}{{Hilton} J}, \bibinfo{author}{{Kopp} G}, \bibinfo{author}{{Kostov} V}, \bibinfo{author}{{Kurtz} DW}, \bibinfo{author}{{Laskar} J}, \bibinfo{author}{{Mason} BD}, \bibinfo{author}{{Milone} EF}, \bibinfo{author}{{Montgomery} M}, \bibinfo{author}{{Richards} M}, \bibinfo{author}{{Schmutz} W}, \bibinfo{author}{{Schou} J} and  \bibinfo{author}{{Stewart} SG} (\bibinfo{year}{2016}), \bibinfo{month}{Aug.}
\bibinfo{title}{{Nominal Values for Selected Solar and Planetary Quantities: IAU 2015 Resolution B3}}.
\bibinfo{journal}{{\em \aj}} \bibinfo{volume}{152} (\bibinfo{number}{2}), \bibinfo{eid}{41}. \bibinfo{doi}{\doi{10.3847/0004-6256/152/2/41}}.
\eprint{1605.09788}.

\bibtype{Article}%
\bibitem[{Qiu} et al.(2021)]{Qiu21}
\bibinfo{author}{{Qiu} D}, \bibinfo{author}{{Tian} HJ}, \bibinfo{author}{{Wang} XD}, \bibinfo{author}{{Nie} JL}, \bibinfo{author}{{von Hippel} T}, \bibinfo{author}{{Liu} GC}, \bibinfo{author}{{Fouesneau} M} and  \bibinfo{author}{{Rix} HW} (\bibinfo{year}{2021}), \bibinfo{month}{Apr.}
\bibinfo{title}{{Precise Ages of Field Stars from White Dwarf Companions in Gaia DR2}}.
\bibinfo{journal}{{\em \apjs}} \bibinfo{volume}{253} (\bibinfo{number}{2}), \bibinfo{eid}{58}. \bibinfo{doi}{\doi{10.3847/1538-4365/abe468}}.
\eprint{2012.04890}.

\bibtype{Article}%
\bibitem[{Raffelt} and {Weiss}(1995)]{Raffelt1995}
\bibinfo{author}{{Raffelt} G} and  \bibinfo{author}{{Weiss} A} (\bibinfo{year}{1995}), \bibinfo{month}{Feb.}
\bibinfo{title}{{Red giant bound on the axion-electron coupling reexamined}}.
\bibinfo{journal}{{\em \prd}} \bibinfo{volume}{51} (\bibinfo{number}{4}): \bibinfo{pages}{1495--1498}. \bibinfo{doi}{\doi{10.1103/PhysRevD.51.1495}}.
\eprint{hep-ph/9410205}.

\bibtype{Article}%
\bibitem[{Reimers}(1975)]{Reimers75}
\bibinfo{author}{{Reimers} D} (\bibinfo{year}{1975}), \bibinfo{month}{Jan.}
\bibinfo{title}{{Circumstellar absorption lines and mass loss from red giants.}}
\bibinfo{journal}{{\em Memoires of the Societe Royale des Sciences de Liege}} \bibinfo{volume}{8}: \bibinfo{pages}{369--382}.

\bibtype{Article}%
\bibitem[{Reimers}(1977)]{Reimers77}
\bibinfo{author}{{Reimers} D} (\bibinfo{year}{1977}), \bibinfo{month}{Oct.}
\bibinfo{title}{{On the absolute scale of mass-loss in red giants. I. Circumstellar absorption lines in the spectrum of the visual companion of alpha $^{1}$Her.}}
\bibinfo{journal}{{\em \aap}} \bibinfo{volume}{61}: \bibinfo{pages}{217--224}.

\bibtype{Article}%
\bibitem[{Ripepi} et al.(2019)]{Ripepi2019}
\bibinfo{author}{{Ripepi} V}, \bibinfo{author}{{Molinaro} R}, \bibinfo{author}{{Musella} I}, \bibinfo{author}{{Marconi} M}, \bibinfo{author}{{Leccia} S} and  \bibinfo{author}{{Eyer} L} (\bibinfo{year}{2019}), \bibinfo{month}{May}.
\bibinfo{title}{{Reclassification of Cepheids in the Gaia Data Release 2. Period-luminosity and period-Wesenheit relations in the Gaia passbands}}.
\bibinfo{journal}{{\em \aap}} \bibinfo{volume}{625}, \bibinfo{eid}{A14}. \bibinfo{doi}{\doi{10.1051/0004-6361/201834506}}.
\eprint{1810.10486}.

\bibtype{Article}%
\bibitem[{Rosseland}(1924)]{Rosseland1924}
\bibinfo{author}{{Rosseland} S} (\bibinfo{year}{1924}), \bibinfo{month}{May}.
\bibinfo{title}{{Note on the absorption of radiation within a star}}.
\bibinfo{journal}{{\em \mnras}} \bibinfo{volume}{84}: \bibinfo{pages}{525--528}. \bibinfo{doi}{\doi{10.1093/mnras/84.7.525}}.

\bibtype{Article}%
\bibitem[{Russell}(1914)]{Russell1914}
\bibinfo{author}{{Russell} HN} (\bibinfo{year}{1914}), \bibinfo{month}{May}.
\bibinfo{title}{{Relations Between the Spectra and Other Characteristics of the Stars}}.
\bibinfo{journal}{{\em Popular Astronomy}} \bibinfo{volume}{22}: \bibinfo{pages}{275--294}.

\bibtype{Article}%
\bibitem[{Salaris} et al.(2022)]{salaris_basti22}
\bibinfo{author}{{Salaris} M}, \bibinfo{author}{{Cassisi} S}, \bibinfo{author}{{Pietrinferni} A} and  \bibinfo{author}{{Hidalgo} S} (\bibinfo{year}{2022}), \bibinfo{month}{Feb.}
\bibinfo{title}{{The updated BASTI stellar evolution models and isochrones - III. White dwarfs}}.
\bibinfo{journal}{{\em \mnras}} \bibinfo{volume}{509} (\bibinfo{number}{4}): \bibinfo{pages}{5197--5208}. \bibinfo{doi}{\doi{10.1093/mnras/stab3359}}.
\eprint{2111.09285}.

\bibtype{Article}%
\bibitem[{Salaris} et al.(2024)]{salaris_24}
\bibinfo{author}{{Salaris} M}, \bibinfo{author}{{Blouin} S}, \bibinfo{author}{{Cassisi} S} and  \bibinfo{author}{{Bedin} LR} (\bibinfo{year}{2024}), \bibinfo{month}{Mar.}
\bibinfo{title}{{Ne22 distillation and the cooling sequence of the old metal-rich open cluster NGC 6791}}.
\bibinfo{journal}{{\em arXiv e-prints}} , \bibinfo{eid}{arXiv:2403.02790}\bibinfo{doi}{\doi{10.48550/arXiv.2403.02790}}.
\eprint{2403.02790}.

\bibtype{Article}%
\bibitem[{Salpeter}(1954)]{Salpeter54}
\bibinfo{author}{{Salpeter} EE} (\bibinfo{year}{1954}), \bibinfo{month}{Sep.}
\bibinfo{title}{{Electrons Screening and Thermonuclear Reactions}}.
\bibinfo{journal}{{\em Australian Journal of Physics}} \bibinfo{volume}{7}: \bibinfo{pages}{373}. \bibinfo{doi}{\doi{10.1071/PH540373}}.

\bibtype{Article}%
\bibitem[{Saumon} et al.(2022)]{saumon22}
\bibinfo{author}{{Saumon} D}, \bibinfo{author}{{Blouin} S} and  \bibinfo{author}{{Tremblay} PE} (\bibinfo{year}{2022}), \bibinfo{month}{Nov.}
\bibinfo{title}{{Current challenges in the physics of white dwarf stars}}.
\bibinfo{journal}{{\em \physrep}} \bibinfo{volume}{988}: \bibinfo{pages}{1--63}. \bibinfo{doi}{\doi{10.1016/j.physrep.2022.09.001}}.
\eprint{2209.02846}.

\bibtype{Article}%
\bibitem[{Schr{\"o}der} and {Cuntz}(2005)]{Schroder05}
\bibinfo{author}{{Schr{\"o}der} KP} and  \bibinfo{author}{{Cuntz} M} (\bibinfo{year}{2005}), \bibinfo{month}{Sep.}
\bibinfo{title}{{A New Version of Reimers' Law of Mass Loss Based on a Physical Approach}}.
\bibinfo{journal}{{\em \apjl}} \bibinfo{volume}{630} (\bibinfo{number}{1}): \bibinfo{pages}{L73--L76}. \bibinfo{doi}{\doi{10.1086/491579}}.
\eprint{astro-ph/0507598}.

\bibtype{Article}%
\bibitem[{Scuflaire} et al.(2008)]{Scuflaire2008}
\bibinfo{author}{{Scuflaire} R}, \bibinfo{author}{{Th{\'e}ado} S}, \bibinfo{author}{{Montalb{\'a}n} J}, \bibinfo{author}{{Miglio} A}, \bibinfo{author}{{Bourge} PO}, \bibinfo{author}{{Godart} M}, \bibinfo{author}{{Thoul} A} and  \bibinfo{author}{{Noels} A} (\bibinfo{year}{2008}), \bibinfo{month}{Aug.}
\bibinfo{title}{{CL{\'E}S, Code Li{\'e}geois d'{\'E}volution Stellaire}}.
\bibinfo{journal}{{\em \apss}} \bibinfo{volume}{316} (\bibinfo{number}{1-4}): \bibinfo{pages}{83--91}. \bibinfo{doi}{\doi{10.1007/s10509-007-9650-1}}.
\eprint{0712.3471}.

\bibtype{Article}%
\bibitem[{Seaton}(2005)]{Seaton2005}
\bibinfo{author}{{Seaton} MJ} (\bibinfo{year}{2005}), \bibinfo{month}{Sep.}
\bibinfo{title}{{Opacity Project data on CD for mean opacities and radiative accelerations}}.
\bibinfo{journal}{{\em \mnras}} \bibinfo{volume}{362} (\bibinfo{number}{1}): \bibinfo{pages}{L1--L3}. \bibinfo{doi}{\doi{10.1111/j.1365-2966.2005.00019.x}}.
\eprint{astro-ph/0411010}.

\bibtype{Article}%
\bibitem[{Siess}(2007)]{Siess07}
\bibinfo{author}{{Siess} L} (\bibinfo{year}{2007}), \bibinfo{month}{Dec.}
\bibinfo{title}{{Evolution of massive AGB stars. II. model properties at non-solar metallicity and the fate of Super-AGB stars}}.
\bibinfo{journal}{{\em \aap}} \bibinfo{volume}{476} (\bibinfo{number}{2}): \bibinfo{pages}{893--909}. \bibinfo{doi}{\doi{10.1051/0004-6361:20078132}}.

\bibtype{Article}%
\bibitem[{Smalley}(2005)]{Smalley05}
\bibinfo{author}{{Smalley} B} (\bibinfo{year}{2005}), \bibinfo{month}{Jan.}
\bibinfo{title}{{T$_{eff}$ and log g determinations.}}
\bibinfo{journal}{{\em Memorie della Societa Astronomica Italiana Supplementi}} \bibinfo{volume}{8}: \bibinfo{pages}{130}. \bibinfo{doi}{\doi{10.48550/arXiv.astro-ph/0509535}}.
\eprint{astro-ph/0509535}.

\bibtype{Article}%
\bibitem[{Straniero} et al.(2020)]{Straniero2020}
\bibinfo{author}{{Straniero} O}, \bibinfo{author}{{Pallanca} C}, \bibinfo{author}{{Dalessandro} E}, \bibinfo{author}{{Dom{\'\i}nguez} I}, \bibinfo{author}{{Ferraro} FR}, \bibinfo{author}{{Giannotti} M}, \bibinfo{author}{{Mirizzi} A} and  \bibinfo{author}{{Piersanti} L} (\bibinfo{year}{2020}), \bibinfo{month}{Dec.}
\bibinfo{title}{{The RGB tip of galactic globular clusters and the revision of the axion-electron coupling bound}}.
\bibinfo{journal}{{\em \aap}} \bibinfo{volume}{644}, \bibinfo{eid}{A166}. \bibinfo{doi}{\doi{10.1051/0004-6361/202038775}}.
\eprint{2010.03833}.

\bibtype{Article}%
\bibitem[{Tononi} et al.(2019)]{tononi2019_wdpopage}
\bibinfo{author}{{Tononi} J}, \bibinfo{author}{{Torres} S}, \bibinfo{author}{{Garc{\'\i}a-Berro} E}, \bibinfo{author}{{Camisassa} ME}, \bibinfo{author}{{Althaus} LG} and  \bibinfo{author}{{Rebassa-Mansergas} A} (\bibinfo{year}{2019}), \bibinfo{month}{Aug.}
\bibinfo{title}{{Effects of $^{22}$Ne sedimentation and metallicity on the local 40 pc white dwarf luminosity function}}.
\bibinfo{journal}{{\em \aap}} \bibinfo{volume}{628}, \bibinfo{eid}{A52}. \bibinfo{doi}{\doi{10.1051/0004-6361/201834267}}.
\eprint{1906.08009}.

\bibtype{Article}%
\bibitem[{Trampedach} et al.(2014)]{Trampedach2014}
\bibinfo{author}{{Trampedach} R}, \bibinfo{author}{{Stein} RF}, \bibinfo{author}{{Christensen-Dalsgaard} J}, \bibinfo{author}{{Nordlund} {\r{A}}} and  \bibinfo{author}{{Asplund} M} (\bibinfo{year}{2014}), \bibinfo{month}{Dec.}
\bibinfo{title}{{Improvements to stellar structure models, based on a grid of 3D convection simulations - II. Calibrating the mixing-length formulation}}.
\bibinfo{journal}{{\em \mnras}} \bibinfo{volume}{445} (\bibinfo{number}{4}): \bibinfo{pages}{4366--4384}. \bibinfo{doi}{\doi{10.1093/mnras/stu2084}}.
\eprint{1410.1559}.

\bibtype{Article}%
\bibitem[{VandenBerg} et al.(2008)]{VandenBerg08}
\bibinfo{author}{{VandenBerg} DA}, \bibinfo{author}{{Edvardsson} B}, \bibinfo{author}{{Eriksson} K} and  \bibinfo{author}{{Gustafsson} B} (\bibinfo{year}{2008}), \bibinfo{month}{Mar.}
\bibinfo{title}{{On the Use of Blanketed Atmospheres as Boundary Conditions for Stellar Evolutionary Models}}.
\bibinfo{journal}{{\em \apj}} \bibinfo{volume}{675} (\bibinfo{number}{1}): \bibinfo{pages}{746--763}. \bibinfo{doi}{\doi{10.1086/521600}}.
\eprint{0708.1188}.

\bibtype{Article}%
\bibitem[{Ventura} et al.(2008)]{Ventura2008}
\bibinfo{author}{{Ventura} P}, \bibinfo{author}{{D'Antona} F} and  \bibinfo{author}{{Mazzitelli} I} (\bibinfo{year}{2008}), \bibinfo{month}{Aug.}
\bibinfo{title}{{The ATON 3.1 stellar evolutionary code. A version for asteroseismology}}.
\bibinfo{journal}{{\em \apss}} \bibinfo{volume}{316} (\bibinfo{number}{1-4}): \bibinfo{pages}{93--98}. \bibinfo{doi}{\doi{10.1007/s10509-007-9672-8}}.

\bibtype{Article}%
\bibitem[{Vink}(2022)]{Vink2022}
\bibinfo{author}{{Vink} JS} (\bibinfo{year}{2022}), \bibinfo{month}{Aug.}
\bibinfo{title}{{Theory and Diagnostics of Hot Star Mass Loss}}.
\bibinfo{journal}{{\em \araa}} \bibinfo{volume}{60}: \bibinfo{pages}{203--246}. \bibinfo{doi}{\doi{10.1146/annurev-astro-052920-094949}}.
\eprint{2109.08164}.

\bibtype{Article}%
\bibitem[{Weaver} et al.(1978)]{Weaver1978}
\bibinfo{author}{{Weaver} TA}, \bibinfo{author}{{Zimmerman} GB} and  \bibinfo{author}{{Woosley} SE} (\bibinfo{year}{1978}), \bibinfo{month}{Nov.}
\bibinfo{title}{{Presupernova evolution of massive stars.}}
\bibinfo{journal}{{\em \apj}} \bibinfo{volume}{225}: \bibinfo{pages}{1021--1029}. \bibinfo{doi}{\doi{10.1086/156569}}.

\bibtype{Article}%
\bibitem[{Weiss} and {Schlattl}(2008)]{Weiss2008}
\bibinfo{author}{{Weiss} A} and  \bibinfo{author}{{Schlattl} H} (\bibinfo{year}{2008}), \bibinfo{month}{Aug.}
\bibinfo{title}{{GARSTEC{\textemdash}the Garching Stellar Evolution Code. The direct descendant of the legendary Kippenhahn code}}.
\bibinfo{journal}{{\em \apss}} \bibinfo{volume}{316} (\bibinfo{number}{1-4}): \bibinfo{pages}{99--106}. \bibinfo{doi}{\doi{10.1007/s10509-007-9606-5}}.

\bibtype{Book}%
\bibitem[{Weiss} et al.(2004)]{Weiss2004}
\bibinfo{author}{{Weiss} A}, \bibinfo{author}{{Hillebrandt} W}, \bibinfo{author}{{Thomas} HC} and  \bibinfo{author}{{Ritter} H} (\bibinfo{year}{2004}).
\bibinfo{title}{{Cox and Giuli's Principles of Stellar Structure}}.

\bibtype{Article}%
\bibitem[Xu et al.(2013)]{XU201361}
\bibinfo{author}{Xu Y}, \bibinfo{author}{Takahashi K}, \bibinfo{author}{Goriely S}, \bibinfo{author}{Arnould M}, \bibinfo{author}{Ohta M} and  \bibinfo{author}{Utsunomiya H} (\bibinfo{year}{2013}).
\bibinfo{title}{Nacre ii: an update of the nacre compilation of charged-particle-induced thermonuclear reaction rates for nuclei with mass number a<16}.
\bibinfo{journal}{{\em Nuclear Physics A}} \bibinfo{volume}{918}: \bibinfo{pages}{61--169}.
ISSN \bibinfo{issn}{0375-9474}. \bibinfo{doi}{\doi{https://doi.org/10.1016/j.nuclphysa.2013.09.007}}.
\bibinfo{url}{\url{https://www.sciencedirect.com/science/article/pii/S0375947413007409}}.

\end{thebibliography*}

\end{document}